\documentclass{article}

\usepackage[margin=1.2in]{geometry}
\usepackage{graphicx,xcolor,hyperref,xspace,placeins,multirow,wasysym, amssymb, enumitem}
\usepackage{tikz}
\usepackage[font=small,margin=20pt]{caption}

\renewcommand{\sp}{\mathop{\sf sp}}
\newcommand{\NST}{\mathop{N_{\sf ST}}}
\newcommand{\PKG}{\mathcal{P}_k(G)}
\newcommand{\ReCom}{{\tt ReCom}\xspace } 
\newcommand{\RevReCom}{{\tt RevReCom}\xspace }
\newcommand{\dwass}{d_{\rm Wass}}

\definecolor{applegreen}{rgb}{0.55, 0.71, 0.0}
\definecolor{teal}{rgb}{0.0, 0.5, 0.5}
\definecolor{alizarin}{rgb}{0.82, 0.1, 0.26}
\definecolor{slategray}{rgb}{0.44, 0.5, 0.56}
\definecolor{amber}{rgb}{1.0, 0.75, 0.0}
\definecolor{mikadoyellow}{rgb}{1.0, 0.77, 0.05}
\definecolor{cadmiumgreen}{rgb}{0.0, 0.42, 0.24}
\definecolor{forestgreen}{rgb}{0.13, 0.55, 0.13}
\definecolor{lust}{rgb}{0.9, 0.13, 0.13}
\definecolor{denim}{rgb}{0.08, 0.38, 0.74}
\definecolor{purpleheart}{rgb}{0.41, 0.21, 0.61}
\definecolor{cherryblossompink}{rgb}{1.0, 0.72, 0.77}
\definecolor{darktangerine}{rgb}{1.0, 0.66, 0.07}
\definecolor{bananayellow}{rgb}{1.0, 0.88, 0.21}
\definecolor{lightblue}{rgb}{0.55,0.82,0.77}
\definecolor{lightgray}{rgb}{0.83, 0.83, 0.83}
\definecolor{languidlavender}{rgb}{0.84, 0.79, 0.87}
\definecolor{jasper}{rgb}{0.84, 0.23, 0.24}
\definecolor{tangelo}{rgb}{0.98, 0.3, 0.0}
\definecolor{tearose}{rgb}{0.97, 0.51, 0.47}
\definecolor{royalfuchsia}{rgb}{0.79, 0.17, 0.57}
\definecolor{cinnabar}{rgb}{0.89, 0.26, 0.2}
\definecolor{cinnamon}{rgb}{0.82, 0.41, 0.12}
\definecolor{bblue}{rgb}{0.502, 0.502, 0.969}

\title{Spanning Trees and Redistricting: \\ New Methods for Sampling and Validation}

\author
{Sarah Cannon,$^{1}$ Moon Duchin,$^{2\ast}$ Dana Randall,$^{3}$ Parker Rule$^{4}$\\
\\
\normalsize{$^{1}$Department of Mathematical Sciences, Claremont McKenna College, Claremont, CA 91711} \\
\normalsize{$^{2}$Data Science Institute, University of Chicago, Chicago, IL 60615}\\
\normalsize{$^{3}$School of Computer Science, Georgia Institute of Technology, Atlanta, GA 30332}\\
\normalsize{$^{4}$Tisch College of Civic Life, Tufts University, Medford, MA 02155 USA}\\
\\
\normalsize{$^\ast$To whom correspondence should be addressed; E-mail:  mduchin@uchicago.edu.}
}
\date{\today}

\begin{document} 
\maketitle 

\begin{abstract}
Deciding whether a political districting plan was distorted by a hidden agenda, or whether it dilutes the voting power of some group, requires a neutral baseline for comparison. 
Remarkably, all nine U.S. Supreme Court justices have now signed on to decisions that find that computational methods can provide key evidence. Today, the leading approaches for benchmarking districting plans are based on the use of spanning trees for sampling graph partitions. We present a new {\em reversible recombination} algorithm and rigorously prove its fundamental properties.  Furthermore, we argue for a canonical sampling distribution called the {\em spanning tree distribution} that is well adapted to redistricting and provides a principled foundation for comparing and validating methods. Together with a highly efficient (and open-source) implementation that can generate and handle large datasets, this work provides the most powerful null model to date for the gerrymandering problem, meeting an urgent democratic challenge with sound scientific methodology.
\end{abstract}

\section{Introduction}

\subsection{Redistricting}
Throughout the world, many countries divide their territory into regions that each conduct legislative elections, from the provinces of South Africa to the departments of France to the states of Brazil.  Any update to the boundary lines---{\em redistricting}---can have a significant impact on  representational outcomes.
Gerrymandering, the practice of abusing line-drawing power by boosting representation for a favored group over other priorities, is constantly in American news---and courtrooms.\footnote{It is a singularly acute problem in the United States because lines must be regularly redrawn, and usually by elected officials themselves.  And the constant high-profile cases come because redistricting cases are among the only ones that still receive mandatory Supreme Court review (28 U.S.C. \S1253).}
Yet for many decades, the U.S. Supreme Court has struggled to identify the non-gerrymandered baseline:  How much representation should groups expect from a ``neutral" redistricting process?

Starting around 2013, experts working in U.S. courts of law have presented computational techniques to sample from the (very large) space of plausible districting plans, arguing that a random {\em ensemble} of plans generated without sensitive data (such as partisan or racial data) provides a neutral statistical baseline. 
The underlying idea of comparing a proposed plan to a collection of alternative plans to diagnose the principles of its design is called the {\em ensemble method} for redistricting analysis.

\begin{figure}[htb!]\centering 
\begin{tikzpicture}[scale=.58]



\begin{scope}[xscale=.8,yscale=.2]
\draw[thick] (2.5,0) rectangle (15.5,30);
\node at (13,-2) {Dem seats};
\foreach \x in {4,5,...,9} {
\node[below] at (\x,-0.1) {\Large \x};  
}

\draw[fill=bblue] (3.6,0) rectangle (4.4,.00015534);
\draw[fill=bblue](4.6,0) rectangle (5.4,.13041567);
\draw[fill=bblue] (5.6,0) rectangle (6.4,12.06381201);
\draw[fill=bblue] (6.6,0) rectangle (7.4,25.37370575);
\draw[fill=bblue] (7.6,0) rectangle (8.4,11.24004759);
\draw[fill=bblue] (8.6,0) rectangle (9.4,1.18596448);

\filldraw [amber] (6,12.06+2) ellipse (.4 and 32/20); 
\filldraw [lust] (6,12.06+4) ellipse (.4 and 32/20); 
\filldraw [purpleheart] (8,11.24+2) ellipse (.4 and 32/20); 
\filldraw [forestgreen] (7,25.37+2) ellipse (.4 and 32/20); 

\draw [dashed] (9,0)--(9,30); 
\draw [red,thick] (18*.4965,0)--(18*.4965,30); 
\draw [teal,line width=2] (7.02574,0)--(7.02574,30) node [above] {7.0};
\node at (12.25,25) {\Large \bf Pres16};
\node at (12.25,20) { (49.65\% Dem)};
\end{scope}



\begin{scope}[xscale=.8,yscale=.2,yshift=-39cm]
\draw[thick] (2.5,0) rectangle (15.5,30);
\node at (13,-2) {Dem seats};
\foreach \x in {3,4,...,9} {
\node[below] at (\x,-0.1) {\Large \x};  
}

\draw[fill=bblue] (2.6,0) rectangle (3.4,0.0031);
\draw[fill=bblue] (3.6,0) rectangle (4.4,1.1785);
\draw[fill=bblue](4.6,0) rectangle (5.4,17.4654);
\draw[fill=bblue] (5.6,0) rectangle (6.4,22.6759);
\draw[fill=bblue] (6.6,0) rectangle (7.4,7.9891);
\draw[fill=bblue] (7.6,0) rectangle (8.4,0.6708);
\draw[fill=bblue] (8.6,0) rectangle (9.4,0.0114);

\filldraw [amber] (4,1.1785+2) ellipse (.4 and 32/20); 
\filldraw [lust] (4,1.1785+4) ellipse (.4 and 32/20); 
\filldraw [purpleheart] (5,17.4654+2) ellipse (.4 and 32/20); 
\filldraw [forestgreen] (6,22.6759+2) ellipse (.4 and 32/20); 

\draw [dashed] (9,0)--(9,30); 
\draw [red,thick] (18*.4928,0)--(18*.4928,30); 
\draw [teal,line width=2] (5.7906,0)--(5.7906,30) node [above] {5.8};

\node at (12.25,25) {\Large \bf Sen16};
\node at (12.25,20) {(49.28\% Dem)};

\end{scope}

\begin{scope}[xshift=16.7cm,yshift=-8.3cm]

\node[draw=amber, line width=4] at (0,10.7) {\includegraphics[width=1.7in]{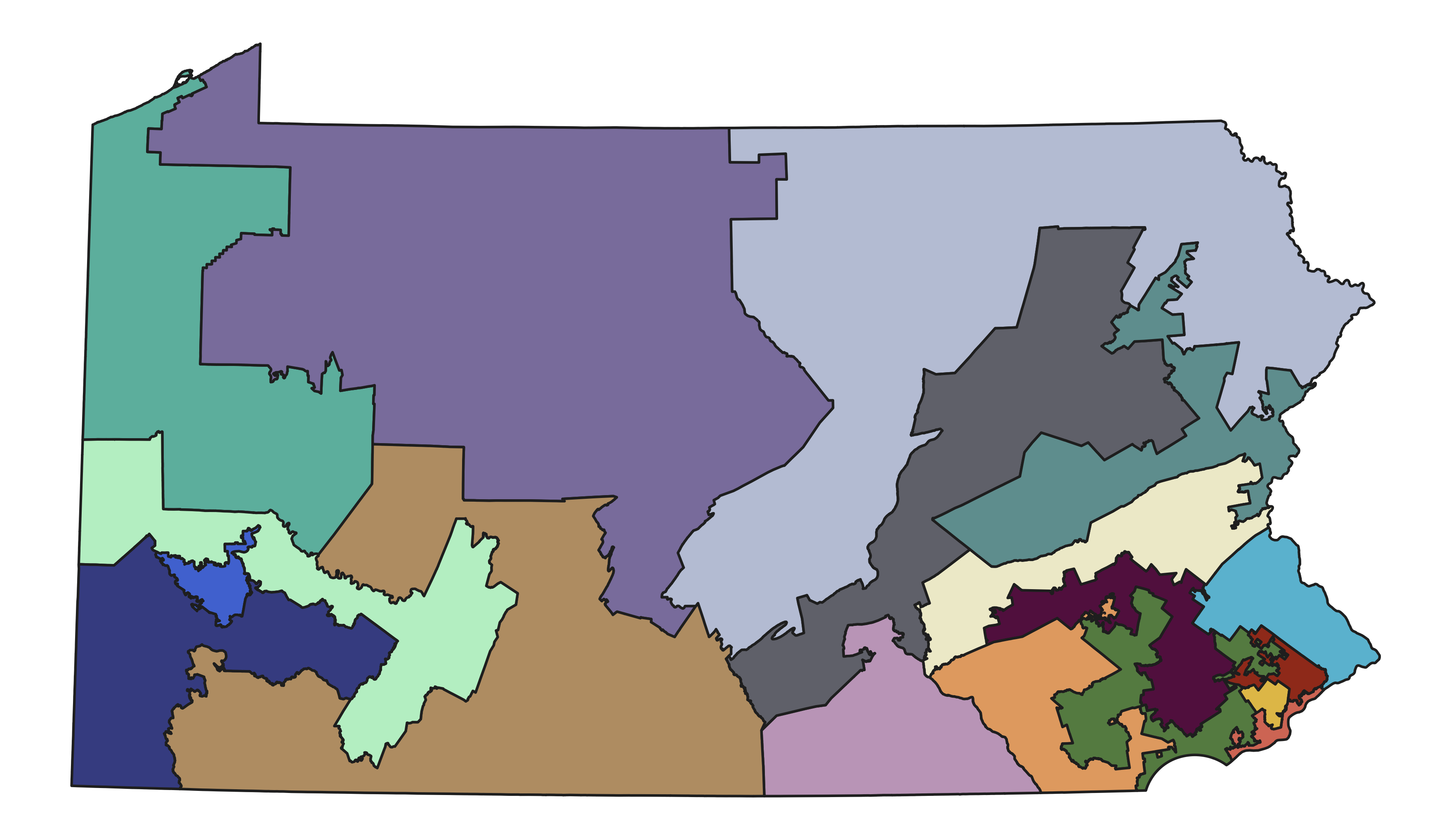}};
\node at (0,7.7) {\large Legislature 2011};
\node[draw=forestgreen, line width=4] at (0,4.1) {\includegraphics[width=1.7in]{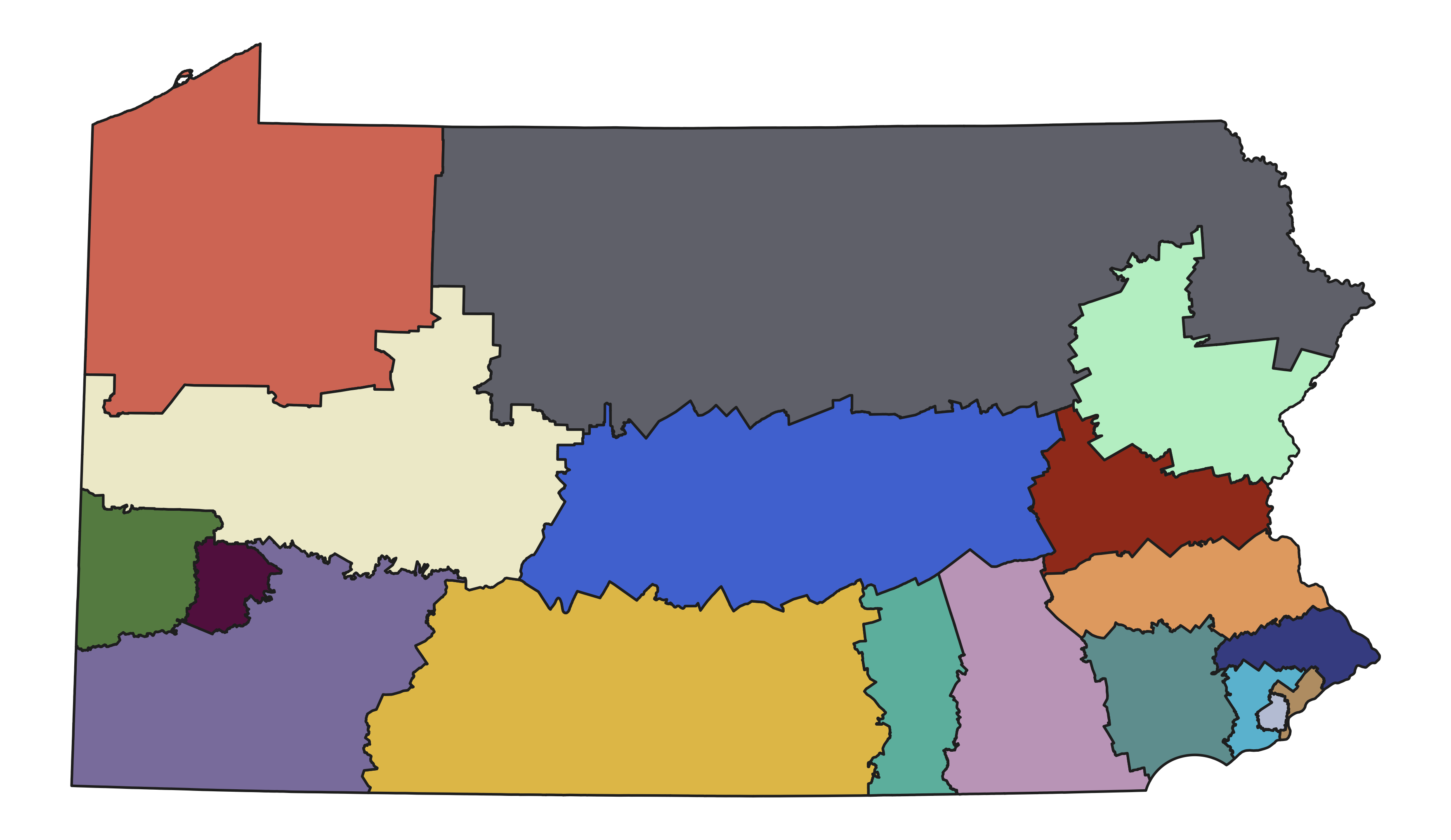}};
\node at (0,1.1) {\large 8th Grade Class};

\node[draw=lust, line width=4] at (8.2,10.7) {\includegraphics[width=1.7in]{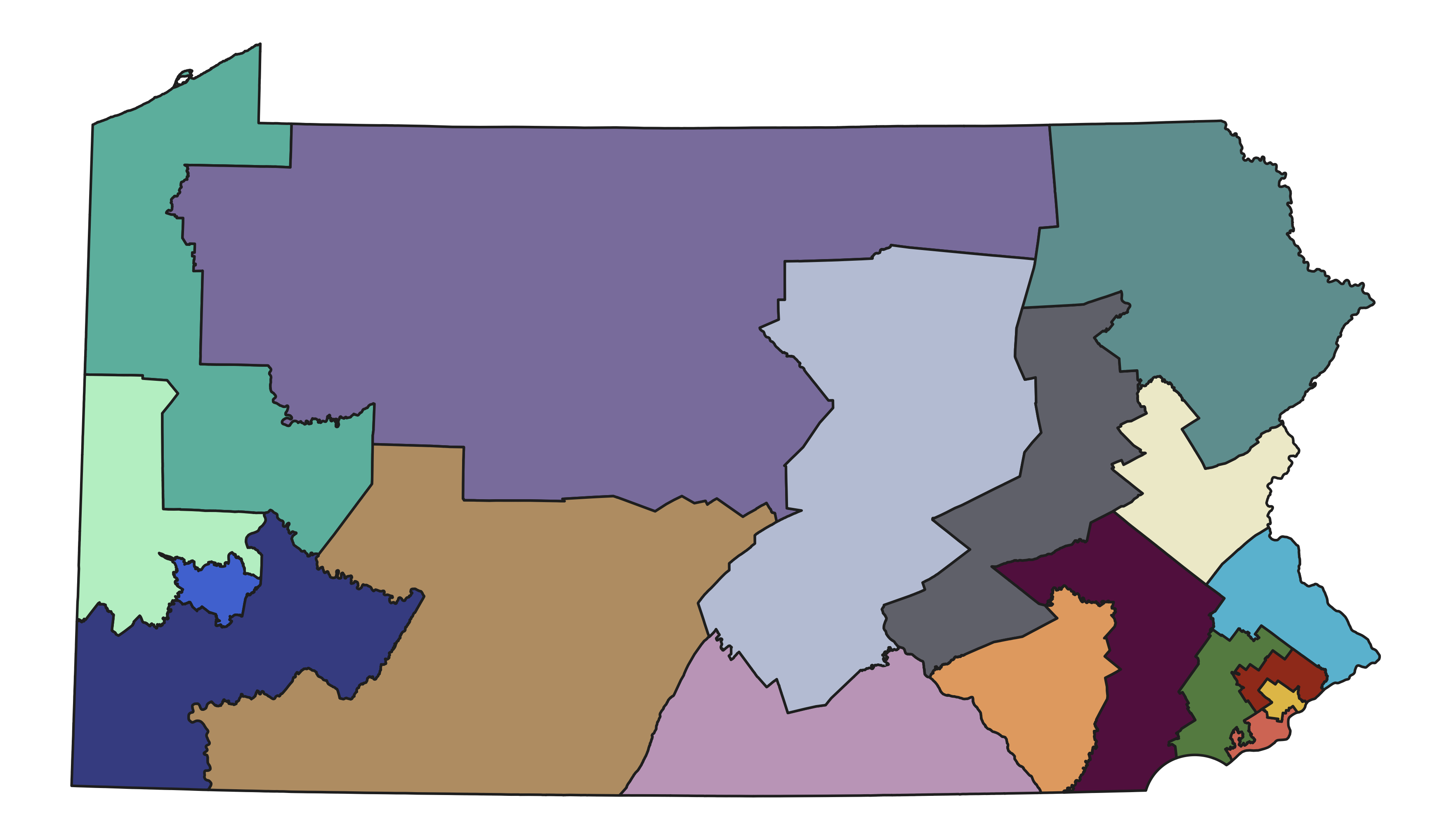}};
\node at (8.2,7.7) {\large Legislature 2018};
\node[draw=purpleheart, line width=4] at (8.2,4.1) {\includegraphics[width=1.7in]{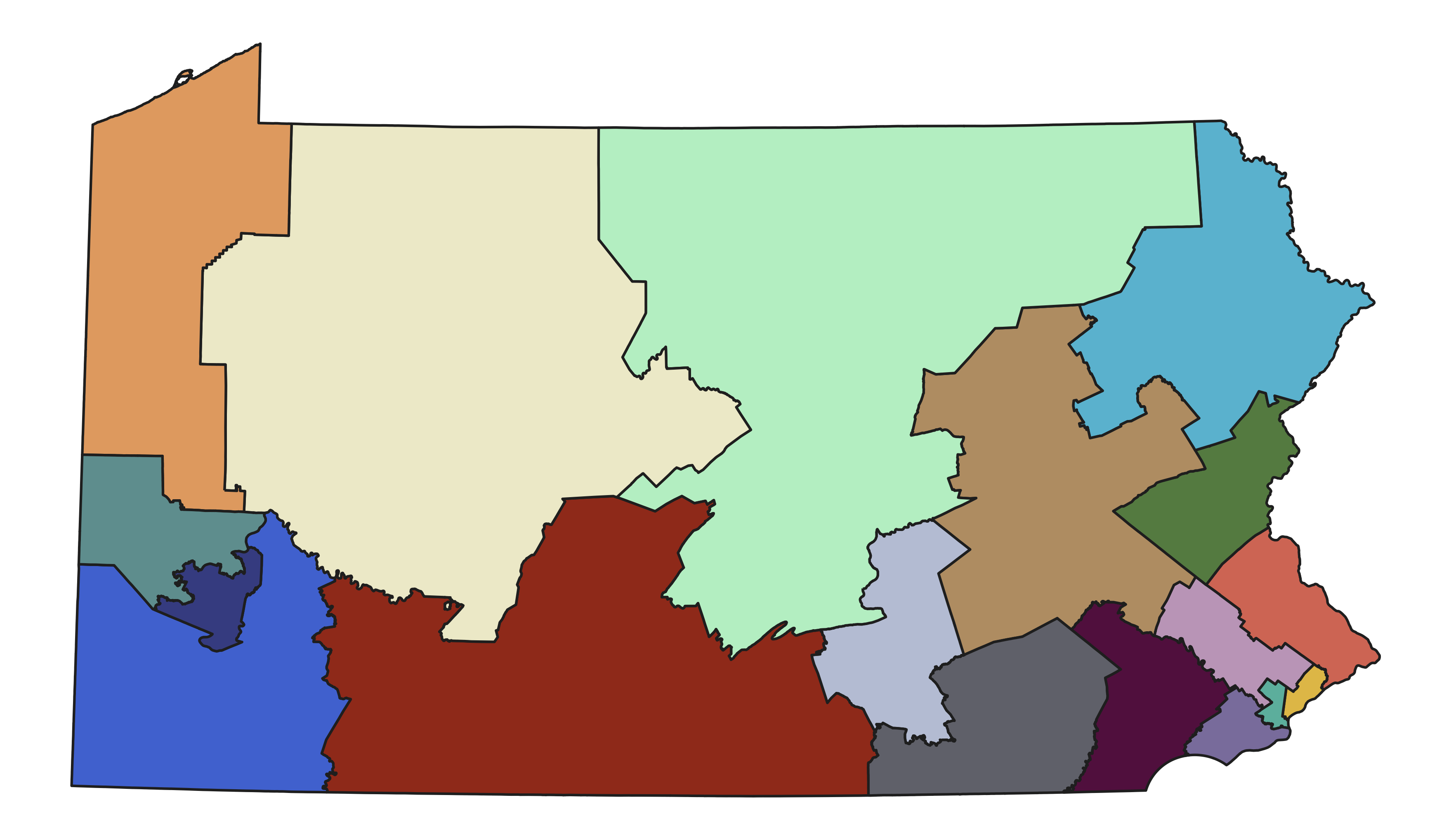}};
\node at (8.2,1.1) {\large Remedial Plan};
\end{scope}

\begin{scope}[xshift=-425pt, yshift=-475pt,xscale=4.7]
    \draw (3.8,.5) rectangle (9.2,5.5);
    \foreach \x in {4,5,6,7,8,9}
    {\draw [gray] (\x,5.5)--(\x,.5) node [below,black] {\Large \x} ;}
    \foreach \x in {5.819,5.728,5.744,5.7685,5.7846,5.786,5.772,5.7885,5.776}
    {\filldraw [fill=gray,opacity=.4,draw=none] (\x-.02,4.5) rectangle (\x+.02,5.5) ;}
\filldraw [fill=teal,opacity=.8,draw=none] (5.7906-.014,4.5) rectangle (5.7906+.014,5.5) ;
\node at ((5.7906,5.5) [above,teal] {5.8};
    \foreach \x in {7.0158,7.0449,7.0158,7.0205,7.0297,7.0157,6.9921,7.1049,7.0712}
    {\filldraw [gray,opacity=.4,draw=none] (\x-.02,4.5) rectangle (\x+.02,5.5) ;}
\filldraw [fill=teal,opacity=.8,draw=none] (7.02574-.014,4.5) rectangle (7.02574+.014,5.5) ;
\node at ((7.02574,5.5) [above,teal] {7.0};
\node at (7.2,5) [right] {\RevReCom};
    
\foreach \x in {7.0237,7.0304,7.0263,7.0203,7.0239,7.0291,7.0513,7.079,7.0532,7.0152}
{\draw [fill=gray,fill opacity=.4,draw=none] (\x-.02,3.5) rectangle (\x+.02,4.5);}
\foreach \x in {5.803,5.7747,5.8187,5.7517,5.8277,5.8064,5.798,5.7974,5.7745,5.773}
{\draw [fill=gray,fill opacity=.4,draw=none] (\x-.02,3.5) rectangle (\x+.02,4.5);}
\node at (7.2,4) [right] {Forest \ReCom};

\foreach \x in {7.0693,6.9549,7.0271,7.1195,7.114,7.0934,7.051,7.0582,7.1269,6.9987}
{\draw [fill=gray,fill opacity=.4,draw=none] (\x-.02,2.5) rectangle (\x+.02,3.5);}
\foreach \x in {5.7819,5.7713,5.8432,5.8153,5.8521,5.7811,5.8059,5.7594,5.8604,5.7383}
{\draw [fill=gray,fill opacity=.4,draw=none] (\x-.02,2.5) rectangle (\x+.02,3.5);}

\node at (7.2,3) [right] {SMC};

\foreach \x in {7.0419,7.035,7.0429,7.0353,7.0372,7.0439,7.0456,7.0452,7.0407,7.0506}
{\draw [fill=gray,fill opacity=.4,draw=none] (\x-.02,1.5) rectangle (\x+.02,2.5);}
\foreach \x in {5.8058,5.8062,5.8012,5.807,5.8084,5.8049,5.8042,5.7971,5.7976,5.802}
{\draw [fill=gray,fill opacity=.4,draw=none] (\x-.02,1.5) rectangle (\x+.02,2.5);}

\node at (7.2,2) [right] {\ReCom-B};

\foreach \x in {7.0388,7.0395,7.0403,7.0395,7.0377,7.037,7.0402,7.0408,7.0393,7.0365}
{\draw [fill=gray,fill opacity=.4,draw=none] (\x-.02,.5) rectangle (\x+.02,1.5);}
\foreach \x in {5.8031,5.8028,5.8047,5.8026,5.8043,5.8044,5.8053,5.8038,5.8056,5.8063}
{\draw [fill=gray,fill opacity=.4,draw=none] (\x-.02,.5) rectangle (\x+.02,1.5);}

\node at (7.2,1) [right] {\ReCom-C};

\node at (7,5.5) [above=11pt] {\large {\bf Pres16} means};
\node at (5.8,5.5) [above=11pt] {\large {\bf Sen16} means};
\end{scope}
\end{tikzpicture}
  \caption{{\bf Illustration of the ensemble method in Pennsylvania.} TOP LEFT: An ensemble of Pennsylvania districting plans produced with \RevReCom is shown in the histograms (in blue). To make it, we overlay the plans with voting from two elections to determine how many of the 18 districts have more D votes than R votes; we call these Democratic seats. These ``blindly drawn" plans usually gives fewer Democratic seats than the proportional outcome of nearly nine---an empirical finding based on the detailed geography of votes. 
TOP RIGHT: Four Congressional plans can then be compared to the ensemble, with colored dots on the histograms marking their performance. The (Republican) legislature's plans both secure more (Republican) partisan advantage than the bulk of the ensemble.
BOTTOM: 
To investigate the consistency of tree-based sampling methods, we repeat this process and plot the mean of each trial.  We include ten independent ensembles made with \RevReCom and with four other samplers
and plot the means with light gray bars, which appear darker when they overlap.  (The means from the histograms above are shown here in teal.)
The methods with asymptotic distributional guarantees (\RevReCom, Forest \ReCom, and SMC) require more computation and give slightly more variable results, while the heuristic \ReCom variants (B,C) give fast and stable results, but come with only an approximate description for their target distribution.  See 
\href{https://github.com/mggg/RRC-Replication/tree/main/figure_and_table_generation/table_outputs}{github} for details.}\label{fig:PA-compare}\label{sec:PA} 
\end{figure}

\clearpage

Ensemble evidence has been used in court cases in at least 11 states since the 2020 Census, and the number continues to grow in the mid-decade redistricting sprint. 
Its wide use in redistricting litigation has led to a series of Supreme Court decisions in which {\em all nine} sitting justices have signed on to opinions describing this kind of evidence as potentially useful for their voting rights work.\footnote{The opinion, concurrences, and dissents in {\em Rucho v. Common Cause} (2019) and  {\em Allen v. Milligan} (2023) contain statements that cover all nine justices---see \S\ref{app:scotus} for more details.}
It is remarkable for a famously math-skeptical Supreme Court to cite a class of graph algorithms as providing useful evidence to address a half-century-old puzzle in their jurisprudence. 

We begin with a motivating example.
Figure~\ref{fig:PA-compare} shows the ensemble method in practice, illustrating issues that arose in {\em League of Women Voters v.~Pennsylvania} in 2018. 
We first demonstrate the method using an ensemble made with \RevReCom, the algorithm introduced here, to randomly draw a large number of districting plans that satisfy population balance, contiguity, and compactness without any partisan data or objectives.
Pairing results from any given election 
with a given set of districts lets us calculate how many seats each political party would win in that vote pattern; these values, across all plans, form histograms like those shown in Figure~\ref{fig:PA-compare}.    Four plans in circulation at the time of the lawsuit are depicted in the figure, with their partisan performance marked in the histograms.  This constructed baseline is precisely what we need to disambiguate geography from gerrymandering in the proposed plans.

These two vote patterns illustrate realistic ways that Pennsylvanians had expressed partisan preference in contemporaneous elections.
Interestingly, although those two elections each had a nearly 50--50 vote split between the major parties, the parties do not tend to get a similar seat split in the random plans:  in the 2016 Presidential vote pattern, Democrats can anticipate controlling 7.0 districts out of 18 in expectation (under 39\%), while the 2016 U.S. Senate vote pattern gives an expectation of just 5.8 seats (roughly 32\%). 
Thus, the use of an ensemble of alternatives lets us measure the effect size of the partisan advantage created by the geography of the voters and the rules of redistricting.
This gap of over a seat in expectation between the two elections is in no way visible by eyeballing a standard map of the voting pattern; it reflects subtle differences in the geography that require an empirical method to draw out.\footnote{In those elections, Donald Trump outpolled Hillary Clinton 2,970,733--2,926,441 in the Presidential contest (Pres16) while the Pat Toomey advantage over Katie McGinty was 2,951,702--2,865,012 in the U.S. Senate race (Sen16)---a nearly equal number of total votes, and a nearly equal split. That means that the difference is not attributable to  ``rolloff" (where one contest had substantially more votes cast) or to third-party candidates, but legitimately reflects the geographical distribution of Clinton voters vs.~McGinty voters.  The analysis reveals, in particular, that there were significant numbers of ticket-splitters (voting R in one contest and D in the other) in both directions.}
For further discussion of the ways that specific geographic configurations can impact representation, see also \cite{GerryIntro,RoddWeigh}.

The Legislature's enacted plan from 2011 (red) and the much more ``compact" proposal from 2018 (yellow) look different to the eye, but they perform similarly against both vote patterns, giving a Republican advantage that pushes beyond the lean of party-neutral plans. The court's remedial plan (purple) is highly responsive to changes in the vote, swinging by three seats under the subtle shift between the two elections. And finally, the plan drawn by an 8th grade class (green) sits at the highest bar both times, behaving just as though it was drawn with no partisan data---which it was.

Figure~\ref{fig:PA-compare}  also lets us compare the seat share estimate from \RevReCom  to values obtained from other methods (introduced below in \S\ref{sec:samplers}) run at their largest practical sample sizes; we see that a  variety of tree-based methods considered in this paper, set to approximately target the same distribution on plans, all produce similar quantitative conclusions about partisan expectation.

\FloatBarrier
\subsection{Graphs, partitions, and Markov chains}

Sampling balanced partitions of a set can be challenging computationally, especially under geometric constraints. In redistricting, not only must
pieces  be connected and population-balanced, but it is also preferable to have compact (nicely shaped) components rather than elongated, spindly, or contorted ones.\footnote{For contorted districts, compare the original gerrymander: \url{https://www.masshist.org/database/1765}.} To make this mathematically
precise, we represent the region by a weighted graph, with nodes for basic indivisible geographical units  and  edges between nodes when the corresponding units are adjacent. The units---such as census blocks or precincts---are weighted by population.

In this formulation, a {\it districting plan} is a partition of the graph into $k$ connected subgraphs with nearly equal population. Building a neutral (non-gerrymandered) baseline for redistricting purposes can be accomplished with an appropriate random sample from the set of balanced graph partitions.  
Setting this up as a graph problem lends itself to a particularly simple measure of compactness via the size of the boundary.

Many methods have been introduced for randomly generating districting plans; for a 60-year overview, see~\cite{GerryAlgs}. 
Here, our main focus will be on Markov chain Monte Carlo (MCMC) algorithms, which are ubiquitous across scientific disciplines for the randomized study of large, complicated configuration spaces (for surveys, see \cite{MCMCRev,AF,Feller}). The idea is to design a random walk that moves among states in a state space---in this case, stepping from partition to partition. Though each plan might only have a few neighbors, MCMC seeks to run until convergence to a useful {\it stationary distribution} (or {\em steady state}) over the entire state space; that is, the probabilities of being at each state will stabilize, so that further steps maintain the probability distribution.

\subsection{Goals}\label{sec:goals}
To create the benchmark for redistricting that the courts have sought, we aim to generate {\em representative samples} of partitions.
This is notably distinct from the frequent use of MCMC for {\em optimization}, where one targets a stationary distribution weighted towards ``better" configurations, seeking local or global optima.  It is also distinct from {\em exploration}, as in generating diverse instances of configurations of an Ising model or spin glass. 
In contrast to these goals, we will designate a meaningful weighting of plans corresponding to a core set of real-world rules and priorities---such as a preference for nicer shapes---and then target this distribution. 
Our ensembles will be built by collecting each plan visited by the random walk; after enough steps,  the statistics of the ensemble should approximate draws from the chain's stationary distribution.\footnote{In many  applications, researchers will employ parameters $s_1$ and $s_2$ to implement {\em burn-in} and {\em sub-sampling}:  a Markov chain process will skip the first $s_1$ states before adding a state to the ensemble; subsequently, every $s_2^{\rm th}$ state visited by the chain will be added. For more discussion of burn-in and sub-sampling, as well as sample size, see \S\ref{sec:burnin_subsample_samplesize}.}
This gives us a notion of typical or expected properties, given a set of rules and priorities.

In this article, we introduce a new {\em reversible recombination} Markov chain (\RevReCom) that provably converges to the precise {\em spanning tree distribution} that various heuristics and algorithms have been built to target.\footnote{We achieve this with combinatorially simple rejection steps rather than a Metropolis filter as in \cite{DukeReCom1}; we will offer empirical comparisons of the approaches, which give mutually reinforcing results.}  
We argue below that this probability measure based on spanning tree counts is an excellent choice of reference distribution for sampling compact graph partitions, well suited to meeting the needs of courts and policymakers.  Finally, we leverage our highly efficient software implementation to expand the repertoire of benchmarking techniques in the literature, offering numerous comparisons of spanning-tree-based samplers and highlighting caveats and limitations. An overarching finding is that, when they are run with adequate sample sizes and clean convergence diagnostics, tree-based samplers provide a growing suite of generally reliable tools.

\FloatBarrier

\section{Partitioning with Spanning Trees}

\subsection{Spanning trees and community structure}\label{sec:spanning}

\begin{figure}[htb!]
\centering
\begin{tikzpicture}
\node at (0,0) {\includegraphics[width=6.5in]{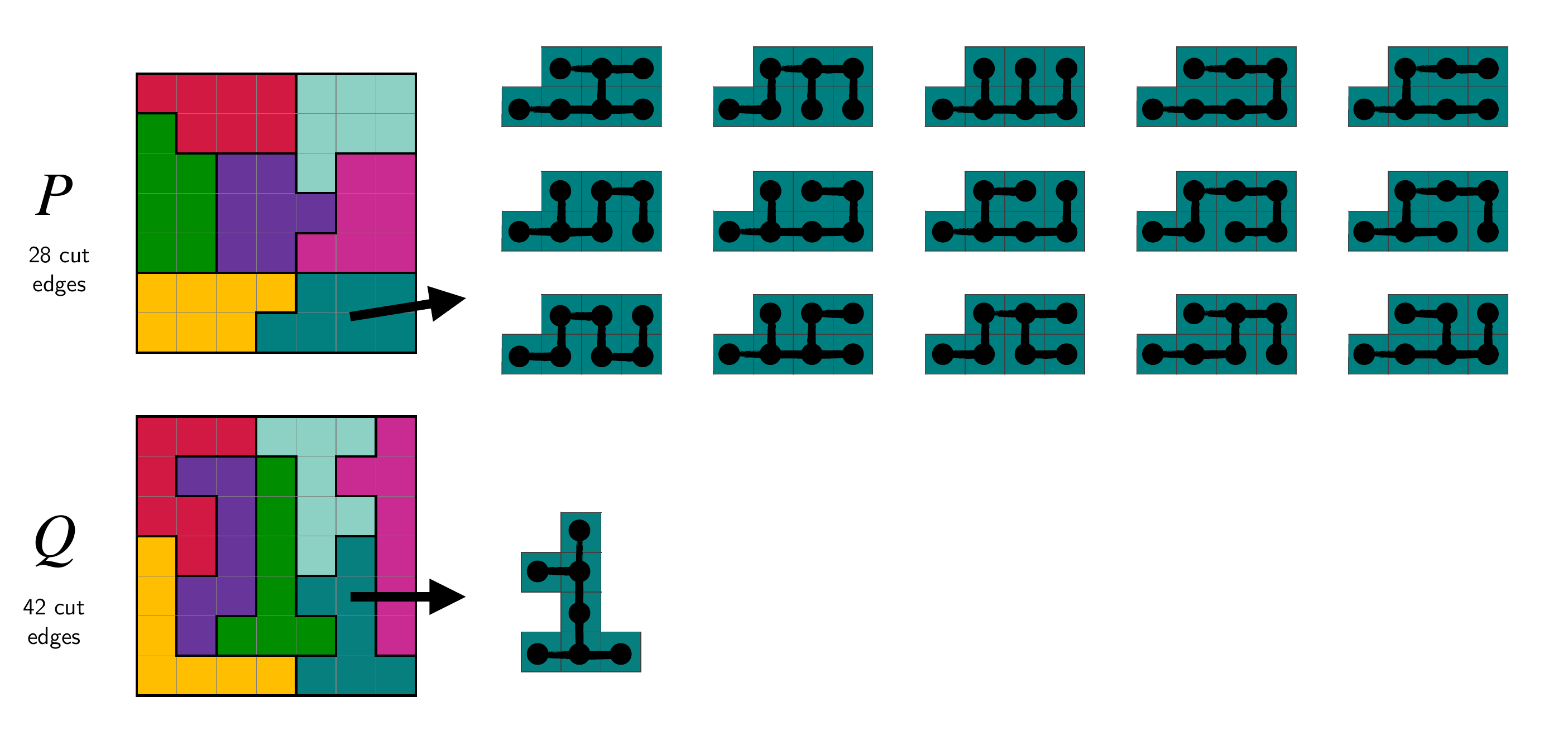}};    
\node at (-5.3,3.8) {\large \bf $7\times 7$ grid, $7$ districts};
\end{tikzpicture}
\caption{{\bf Weighting partitions with spanning trees.}
Two configurations or districting plans $P$ and $Q$ are shown here, for the $7\times 7\to 7$ districting problem. There are 28 cut edges in plan $P$ and 42 cut edges in plan $Q$. In this example, each district $Q_i$ in $Q$ has just one spanning tree, while each district $P_i$ in $P$ has 15 spanning trees. This means a sample from the spanning tree distribution $\pi$ is exactly $15^7$ times as likely to choose the ``plump" plan $P$ as it is to choose the ``spindly" plan $Q$---a factor of over 170 million.}\label{fig:spanning}
\end{figure}

A vast number of applications in theoretical computer science and engineering rely on the use of {\em spanning trees} of a graph $G$, which are cycle-free subgraphs of $G$ using all vertices. 
They contain the minimum amount of connective tissue to keep all the nodes in one component; that is, deleting any edge cuts the graph into (exactly) two pieces.
The number of spanning trees $\NST(G)$ of a graph $G$, sometimes referred to as its {\em complexity}, is a measure of richness or well-connectedness that is efficiently computed using graph Laplacians \cite{Lyons}.  A simple graph on $n$ nodes can have a spanning tree count anywhere from 1 (if it is itself a tree) to $n^{n-2}$ (if all possible edges are present); the number grows quickly as the graph gets larger and more internally connected. Figure~\ref{fig:spanning} highlights two pieces or ``districts" in partitions $P$ and $Q$ of a grid-graph.
One highlighted district has 15 spanning trees and the other has just 1 spanning tree (because it is a tree itself).  
We can take the size of the cut-set of a partition (the number of edges with their endpoints in different parts, also called the number of {\em cut edges}) to be a measure of the efficiency of the partition.  

Small cut-sets, high internal connectivity within districts, and efficient-looking shapes are all simultaneously achieved when the pieces of a graph partition have relatively many spanning trees \cite{DuchinTenner}.  
Accordingly, a natural choice of weighting for a graph partition is to use the product of the spanning tree counts of its districts---in other words, if $P$ is a partition of a graph into connected subgraphs $P_1,\dots,P_k$ representing its districts, then we want to sample according to $\NST(P):=\prod_i \NST(P_i)$.
We thus define a weight proportional to this spanning tree count,
$$\pi(P) \propto \NST(P),$$
and call this the {\em spanning tree distribution} on partitions. 
It is intuitively clear that weighting by $\pi$ favors ``plump" districts, because those typically have more spanning trees than more ``spindly" or tree-like districts do, as illustrated in Figure~\ref{fig:spanning}.

\begin{table}[bht!]\centering

\begin{tabular}{lll}
cut &number of plans&total spanning tree weight\\
edges & (exact count) &  (millions, rounded)\\
\hline 
28&	420&   73447\\
29&	5408&   250666\\
30&	43468&   528671\\
31&	219704&   698394\\
32&	884620&   732191\\
33&	2686928&   577890\\
34&	6578950&   366429\\
35&	12985744&   186993\\
36&	21167576&   78541\\
37&	28289752&   26977\\
38&	31084950&   7589\\
39&	27036848&   1684\\
40&	17848860&   282\\
41&	7971064&   32\\
42&	1949522& 2  
\end{tabular}

\caption{{\bf Spanning tree weight factors.} This table quantifies by how much the spanning tree distribution up-weights more compact plans (those with a smaller cut-set) relative to the complete enumeration---the center column is uniform, while the right column is formed by summing the spanning tree weight $\NST(P)$ over all plans with each number of cut edges. The $L^1$ Wasserstein distance between distributions can be calculated after normalizing each to total mass one, obtaining $\dwass({\rm uniform},\pi)=5.72785$.  The average cut edge count shifts correspondingly, from roughly 37.61 (when all plans are equally weighted) to 31.88 (when we sample from the distribution $\pi$). This represents a marked improvement in compactness.}\label{tab:enumeration}
\end{table}

For the problem of dividing a $7 \times 7$ grid into $7$ districts of equal size, partition $P$ from Figure~\ref{fig:spanning} is $15^7$ times more likely to appear as a sample under this distribution than partition $Q$ is.   This very heavy preference is balanced out by the fact that there are many more plans with long boundaries than there are highly compact plans.
Across all partitions of the $7 \times 7$ grid, 
Table~\ref{tab:enumeration} confirms the overall impact of re-weighting by spanning tree count:  more compact partitions with shorter boundaries are favored in this distribution, bringing the average size of a cut-set down from roughly 37.6 to 31.9.

Conceptually, the spanning tree count reflects how effectively the plan picks out strongly internally connected districts in a geographical network, under the additional constraint of balanced population. 
Selecting highly connected subgraphs  has an ample literature of its own, going by the name {\em community detection} in the network science literature \cite{POM,MooreCommunity}.  The explicit use of spanning tree counts for community detection is employed in numerous papers, including \cite{KW}. 
Whether this application of clustering might correspond well to {\em social} understandings of community is an interesting question that was broached by Nelson in \cite{Nelson}, and is worth serious attention in future work.

There is a growing body of work exploring more precise relationships between spanning tree counts and cut-sets.  Clelland et al.~investigate, for $k=2$ districts, the strong (negative) linear relationship between $|\partial P|$ (the cut edge count) and the modified spanning tree weight $\log\left(\NST(P) |\partial P|\right)$, in both grids and real-world examples~\cite{clelland2021compactness}. 
A subsequent paper by Procaccia and Tucker-Foltz 
\cite{procacciaTuckerFoltz2021compactness}  uses effective resistance to prove an asymptotically exponential relationship between the cut edge count and the spanning tree count. The constants in this exponential relationship can vary depending on the structure of the graph.  In the special case of grid-graphs, the strongest known results are those given by Tapp \cite{Tapp}.

If working within the statistical physics paradigm, one may be tempted to try to sample plans with small values for $|\partial P|$ more directly by targeting a distribution proportional to $e^{-\beta |\partial P|}$. One downside of this approach is the added complication/discretion of choosing the value of $\beta$.  Another issue is that known results show that sampling from this distribution is NP-hard in some planar graphs~\cite{NDS1}.

\FloatBarrier

\subsection{Tree-based sampling methods}\label{sec:samplers}

Earlier research on redistricting with Markov chains focused on local or quasi-local moves that change the district assignment of one or a small number of units; these are generally called {\em flip steps} because they can be visualized as flipping the `spin' or color of the nodes, analogous to Glauber dynamics in statistical physics \cite{GerryAlgs}.
The difficulties of sampling approximately balanced partitions in a flip chain are well known, even on simple grid graphs, as flip chains face extremely slow convergence  (similar to low-temperature models in statistical physics)~\cite{ReCom,frieze2022subexponential,NDS1}. 
In other words, these chains have prohibitively long {\em mixing times}---they require many steps to pass a threshold of closeness to the stationary distribution.
Often, allowing large, non-local perturbations in a single move can significantly speed up a Markov chain, as in various classic card-shuffling examples.\footnote{Using riffle-type moves that affect the whole deck produces significantly faster mixing than iterating single-card moves---the mixing time drops to the order of $\log n$ instead of $n\log n$  for decks of $n$ cards~\cite{AD,DS,diaconis-book,BD}.}  

A  family of large-step Markov chains called {\em recombination} (or \ReCom) was introduced by DeFord, Duchin, Najt, and Solomon \cite{ReCom,NDS1,NDS2}.
The basic recombination step  proposes two districts to be fused, then selects a random spanning tree of the fused double-district, then (if possible) chooses an edge from that spanning tree whose deletion leaves two components with population balance within the prescribed tolerance, thereby defining two new districts.

There are several natural variants within the \ReCom family,
based on how the choices of adjacent districts and spanning trees are randomized. One choice is whether to select the pair of districts to merge by randomizing indices or by taking the districts spanned by a random cut edge of the partition; a second choice is whether to generate a random spanning tree uniformly (UST), or by randomizing edge weights and taking the minimum-weight spanning tree (MST). In the comparisons that follow, we will designate these as \ReCom-A (cut edge, MST), \ReCom-B (district index, MST), \ReCom-C (cut edge, UST), and \ReCom-D (district index, UST).  
Surprisingly, while each of these \ReCom variants approximately targets the spanning tree distribution, no closed-form representation of their precise steady state is known when there are more than two districts.
We compare these variants empirically below.


These ideas---leveraging spanning trees for graph partitioning, and targeting the spanning tree distribution $\pi$ described here---have inspired multiple related samplers.  Mattingly et al. introduced a Metropolis-Hastings variant called Forest \ReCom \cite{DukeReCom1,DukeReCom2}.  Without further tuning (i.e., with parameters $\gamma =\beta = 0$ in their implementation), the resulting process targets $\pi$. Later, Charikar et al.~proposed an alternative method to target $\pi$ in \cite{charikar2022complexity}, incorporating some additional balance factors.  Together with a recent result of Cannon--Pegden--Tucker-Foltz that confirms that a polynomial fraction of trees admit a balanced cut in grid-like graphs \cite{splittable}, the Charikar team's method runs in polynomial time---but still slowly in practice.
Abrishami et al.~have implemented a direct-sampling method to cut a single tree $k-1$ times and obtain a $\pi$-distributed partition.\footnote{Code can be found \href{https://github.com/tabrish/direct-sampling}{here} and \href{https://github.com/inkyubeytor/lupartition}{here}.}  
And McCartan--Imai have developed an importance sampling method called sequential Monte Carlo, or SMC, that works with many copies of the initial graph and draws spanning trees to mark additional districts in a multi-stage process \cite{SMC}.  Here too, the default target is $\pi$.\footnote{This method is quite popular among political scientists.  Interestingly, part of the reason is that it is coded in R rather than Python.}
New methods have been accompanied by a recent string of theoretical advances regarding properties of recombination chains \cite{akitaya2022reconfiguration_recombination,procacciaTuckerFoltz2021compactness,frieze2022subexponential,irred}.   

Here, we introduce a simple modification to \ReCom that makes it {\em reversible}, letting us show that the steady state is precisely the spanning tree distribution on plans.
This is achieved by adding more rejection conditions to make the chain  {\em lazier}, or more likely to self-loop, putting higher multiplicity on certain states in a controlled manner.  We compensate for the high rejection rates with a parallelization strategy described in \S\ref{subsec:implementation}, maintaining high efficiency overall.
With its rigorous foundation together with a fast implementation and customized data pipeline and compression scheme, \RevReCom lets us conduct large experiments and make extensive comparisons between many of the spanning tree methods used in the field. 
Our primary experimental comparisons will put \RevReCom alongside Forest \ReCom, SMC, and the original \ReCom variants A,B,C,D.  Throughout the paper, we ran each method at its {\em largest practical sample size}---this means we used the authors' latest software implementations with uniform resource  constraints of time ($48$ hours of runtime), memory ($100$GB RAM), and  storage ($2$GB after efficient compression).  

\section{Reversible ReCom}
\label{sec:revrecom}


\subsection{Exact balance}\label{subsec:exact}

We first define reversible recombination  (\RevReCom) for the simple case where nodes have equal weight and graph partitions require exact balance; we relax these assumptions in Section~\ref{subsec:approx}. 
Let $\PKG$ be the set of connected $k$-partitions on a graph $G$ with an equal division of nodes. Let $P_1, \dots, P_k$ denote the $k$ subgraphs induced by the pieces of the vertex partition, which we will call {\em districts}.\footnote{To make terminology simpler, we use the term ``districts" to refer either to the subsets of vertices or to the subgraphs on those subsets.} 
We call an edge $e$ of a tree $T$ a {\em balance edge} if its two complementary components have the same number of nodes.
 For subgraphs $A,B$ of $G$, let $E(A,B)$ be the set of edges in $G$ with one endpoint in $V(A)$ 
and one in $V(B)$.  Let $\NST(A,B)$ be the number of spanning trees of the induced graph on $V(A)\cup V(B)$ that have exactly one edge in $E(A,B)$; since such a tree consists of a spanning tree on each of A and B plus that one edge,
$$\NST(A,B)=\NST(A)\cdot \NST(B) \cdot |E(A,B)|.$$

Then the spanning tree distribution $\pi$ on $\PKG$ is defined by
$$\pi(P) := \frac{\prod_{i=1}^k \NST(P_i) }{Z},$$ 
where $Z = \sum\limits_{R\in\PKG} \prod_{i=1}^k \NST(R_i)$ is the {\it normalizing constant} ensuring $\pi$ is a 
distribution.\footnote{We can also define the quantity 
$\sp(P):=\ln\left(\prod_{i=1}^k \NST(P_i)\right)$ as the {\em spanning tree score} of a districting plan. This has been proposed in its own right as a measure of the compactness of a plan, with higher scores indicating greater compactness, as in  \cite{DuchinTenner}.  This lets us write $\pi$ in a format familiar in Markov chain theory as $\pi(P) := \displaystyle\frac{ e^{\sp(P)}}{Z}$.}

We define the Markov chain proposal as follows.  From a plan $P$, the fusion step is made according to district indices:  first choose pairs of indices uniformly from $\{1,\dots,k\}^2$ and let $A=P_i$ and $B=P_j$. If $A$ and $B$ are not adjacent and distinct, reject. 
If they are adjacent and distinct, consider the graph $A\cup B$ induced by $G$
on the vertex set $V(A)\cup V(B)$ and choose a spanning tree uniformly at random.  If there is no balance edge, reject.
If there is, then it is unique; delete the balance edge and let the new components be called $A',B'$ and the corresponding new plan be called $Q$.
Accept $Q$ with probability $1/|E(A',B')|$; else, reject. If the acceptance condition was not met, resample a new pair of indices and repeat. 

Note that the number of spanning trees for $A \cup B$ that could have produced districts $A'$ and $B'$ is exactly $\NST(A', B')$. 
The process above prescribes a  transition probability $X_P(Q)$ for transitioning from state $P$ to state $Q$ in a single step, as follows:
$$X_P(Q)= \frac{2}{k^2} \cdot \frac{1}{\NST(A\cup B)} \cdot \frac 1{E(A',B')} \cdot \NST(A',B')= \frac{2}{k^2}\cdot \frac{\NST(A')\NST(B')}{\NST(A\cup B)}$$
if suitable $A,B,A',B'$ exist (in which case they are determined up to relabeling), and zero if not.  
Noting that $ V(A)\cup  V(B)=V(A')\cup V(B')$ and also that $P_\ell = Q_\ell$ for $\ell \neq i,j$, one can now verify:
\[
\pi(P)\cdot X_P(Q)  
= \frac{2}{k^2 Z} 
\frac {\prod\limits_{\ell \neq i,j} \NST(P_\ell)}{\NST(A\cup B)} \NST(A)\NST(B)\NST(A')\NST(B') 
= \pi(Q) \cdot X_Q(P)
\]
The condition that 
$\pi(P)\cdot X_P(Q)= \pi(Q)\cdot X_Q(P)$ is called {\em detailed balance}, and when it holds for all $P,Q$, the Markov chain is said to be {\em reversible}.  Satisfying detailed balance with $\pi$ ensures that it is an equilibrium distribution because if we start with probabilities distributed by $\pi$, the probability of being at any state $P$ after one application of $X$ is given by $\sum_Q \pi(Q) \cdot X_Q(P) = \sum_Q \pi(P) \cdot X_P(Q) = \pi(P)$. 

\begin{figure}[htb!]
\centering
\includegraphics[width=6.5in]{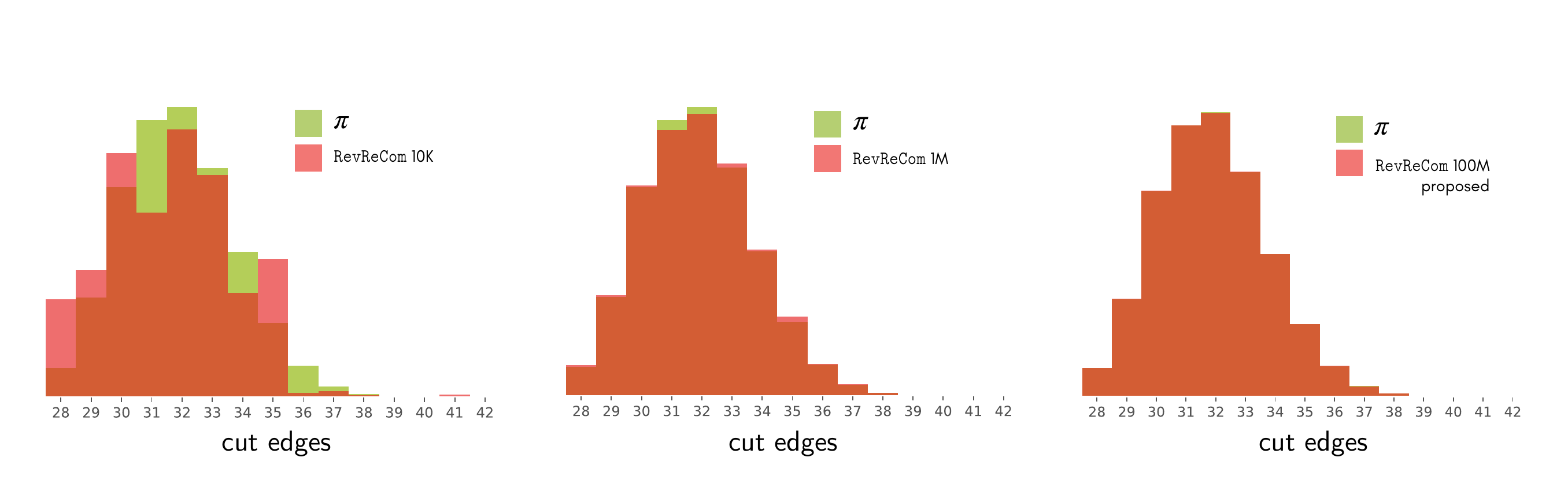}
\caption{{\bf Convergence in distribution.} For the $7 \times 7$ grid divided into $7$ equal-sized districts, these plots show the distribution of cut edges in an ensemble of districting plans created after 10,000, 1 million, and 100 million \RevReCom  proposal steps (red) compared to the distribution of cut edges in all districting plans, weighted by  $\pi$ (green). 
The million-step sample was collected in under ten seconds on a laptop computer, and the time growth is linear.}\label{fig:rrc-convergence}
\end{figure}

An example comparing the cut-edge distribution of samples created with \RevReCom with 10,000, 1 million, and 100 million steps to the cut-edge distribution of the true underlying distribution $\pi$ can be see in Figure~\ref{fig:rrc-convergence}.

\subsection{Approximate balance}\label{subsec:approx}

Next, we treat the case where we only require approximate balance for the populations across districts, rather than exact balance as on grid-graph examples. Real-world examples call for this relaxed approximate-balance formulation.

Given population weights on the vertices $w:V\to \mathbb R$ and 
a small population deviation tolerance $\epsilon>0$, a graph partition $P$ into $k$ districts is $\epsilon$-{\it approximately balanced} if for all districts $P_i$,  
$$(1-\epsilon) \frac{w(G)}k\le w(P_i) \le (1+\epsilon) \frac{w(G)}k,$$
where $w(G) = \sum_{v \in V(G)} w(v)$.

In this setting, a tree may have multiple edges whose removal separates the vertices into two approximately balanced parts. 
We call an edge $e$ of a tree $T$ an {\em $\epsilon$-balance edge} if its two complementary components have population balance within tolerance $\epsilon$; when $\epsilon = 0$, this is exactly the notion of a balance edge introduced in the main text.

We modify the algorithm as follows.  Let $m$ be a global upper bound on the number of $\epsilon$-balance edges that can exist in any tree spanning a double-district-sized subgraph of $G$.  
We will accept a candidate balance edge with probability $1/m$.  If there are $b$ balance edges in a particular spanning tree $T$, this means there is a probability $1-\frac bm$ that no edge will be chosen, and the proposal will be rejected at this stage. 
Recalculating the transition probability:   
$$X_P(Q)= \frac{2}{k^2} \left(\frac{\NST(A',B')}{\NST(A\cup B)}\right)\cdot \frac 1{m\cdot E(A',B')}= \frac{2}{m k^2}\cdot \frac{\NST(A')\NST(B')}{\NST(A\cup B)},$$
and detailed balance follows as before.  Because of the rejection probability of $1 - \frac bm$, it is advantageous to find an upper bound $m$ that is close to tight. Certainly one can set $m$ to the total number of edges, but this unnecessarily shrinks the acceptance probability.  In practice, we have chosen a large value of $m$, tolerating a high rejection rate in order to be more confident that our bound is globally valid.  In our Pennsylvania and Virginia runs, for example, the mean value of $m/b$ (where $b$ is the actual observed number of balance edges in a given proposal) is between four and five. 

If the chosen  $m$ is not a global upper bound, the process may not converge to $\pi$. Even if the number of balance edges is never observed to exceed $m$ in a given run, the resulting limiting distribution is conditioned on never encountering a spanning tree that has more than $m$ balance edges. This could differ from $\pi$ in subtle ways, an issue that equally impacts \RevReCom and the SMC sampler.

To guarantee the existence of unique steady state that attracts every starting distribution for a Markov chain, we must verify that the chain is ergodic, i.e., aperiodic and irreducible.  In general, irreducibility for recombination chains is a hard open problem, though there is a great deal of recent work in that direction.  
See \S\ref{subsec:obstruc-continued} for an extended discussion of ergodicity. 

\section{Empirical Results}
\label{sec:results}

\subsection{Benchmarking with summary statistics}
Quantities of interest assessed for a plan can be thought of as a projection from the state space to $\mathbb R^n$, or just to the real numbers in the case of a single metric.  These projections are of interest both because summary statistics may capture many of the relevant features of a plan (like its partisan performance, its number of county splits, and so on), and because lower-dimensional representations can be easier to visualize and to understand.  

In the demonstrations here we use the push-forward to  plan-wide numerical scores to present empirical results.  Some scores are shape-based, like the cut edge count (Figures~\ref{fig:rrc-convergence} above and \ref{fig:7x7} below).  Some are partisan, like the count of districts favoring some political party (Figure~\ref{fig:PA-compare}). In addition to plan-wide scores, we will also study  district-level numerical scores like the partisan vote share by district (Figure~\ref{fig:VA}).

It is easy to see that convergence in the push-forward distribution to a summary statistic may occur well before representative sampling in the state space overall.
This means that the quality of convergence in a projected statistic only gives an inequality bound on convergence in the full state space---failure to converge in projection ensures failure to converge overall, but success provides no guarantees.  Nevertheless, using a statistic that varies in a way that is related to the stationary distribution  will clearly give a more discerning view of convergence than one that is likely to have less or no relationship to $\pi$.  For this reason, when sampling from the spanning tree distribution, the cut edge plots will be far more likely to catch convergence problems than the partisan plots.    

To assess the similarity of sampling distributions, we will use 
{\em Wasserstein distance} (also known as  earth-mover distance), which may be familiar to  readers from optimal transport theory.  
The Wasserstein distance between two distributions over a common metric measure space computes the cost of shifting the probability mass from one distribution to the other.\footnote{For instance, if a histogram with total mass one is moved by adding six to every recorded value, then the total cost of the move is 6---here, implicitly, neighboring bins are regarded as one unit apart.  If some transport plan moves 15\% of the mass of a distribution to a location two units away from its start, then its total cost is $0.3$.}  The Wasserstein distance between two distributions is given by the lowest-cost transport plan; for single quantities, this is the integral of the absolute difference between the two cumulative distribution functions (CDFs), meaning that there is no need to identify a minimum-cost matching to compute the distance. 

In Markov chain theory, mixing time is customarily measured with total variation distance, which is a normalization of $L^1$ distance between probability distributions on the state space.  Convergence in total variation distance implies convergence in ($L^1$) Wasserstein distance, and Wasserstein distance on projected statistics is used here.

\FloatBarrier
\subsection{Benchmarking against a complete enumeration}\label{sec:7x7}

In most practical applications, it is not feasible to give precise bounds on mixing time; instead, practitioners rely on established convergence diagnostics to give heuristic evidence that results are reliable. These include {\em multi-start tests}, where chains are run from different starting configurations to be sure they are escaping the neighborhood where they were initialized.  We also perform {\em enlargement tests}, collecting samples long past the point that statistics appear to stabilize in order to create more opportunity to identify bottlenecks causing pseudo-convergence.

\begin{figure}[htb!]\centering
\begin{tikzpicture}
\node at (0,8){\includegraphics[width=6.2in]{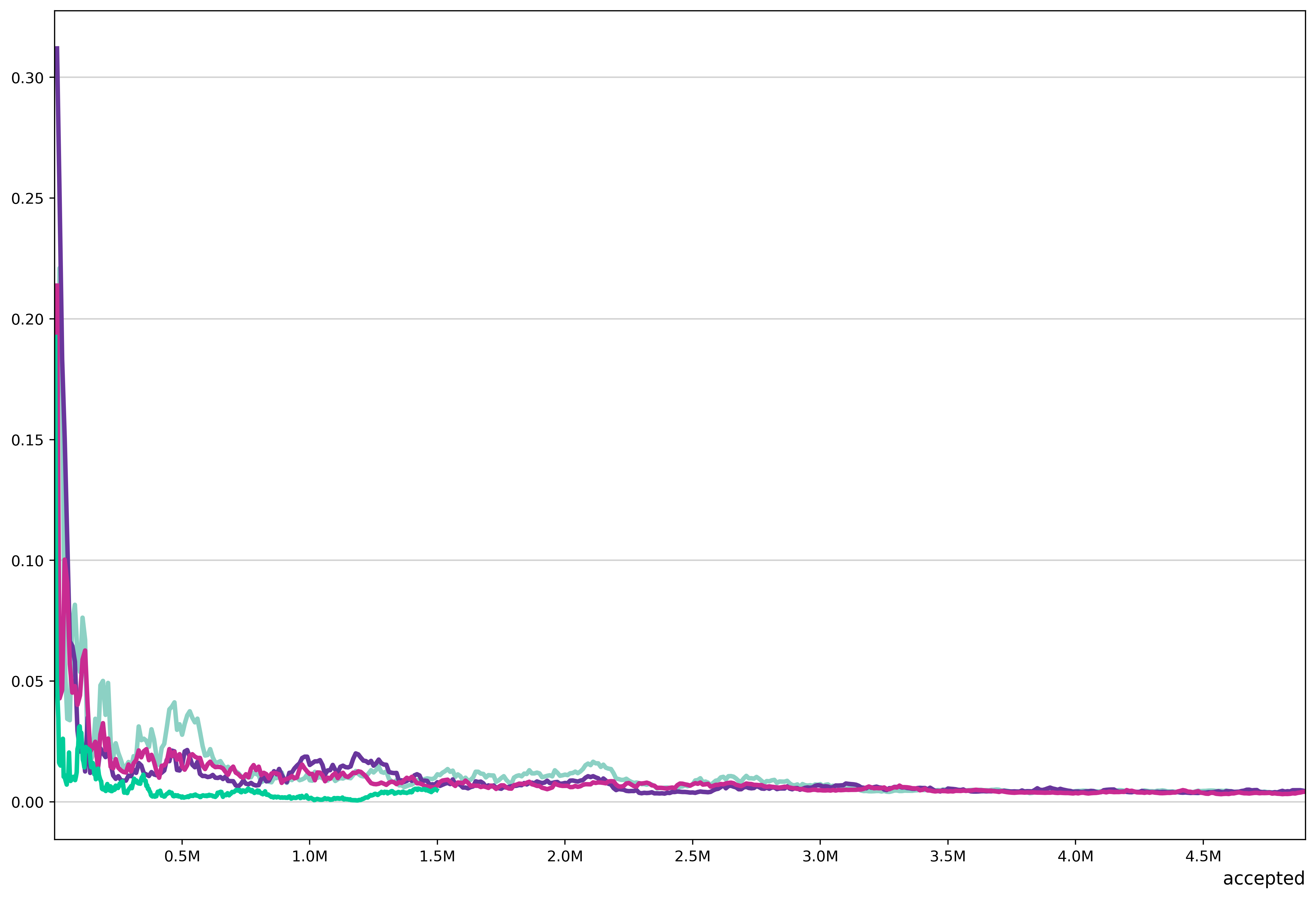}};
\node at (5,12){\includegraphics[width=140pt]{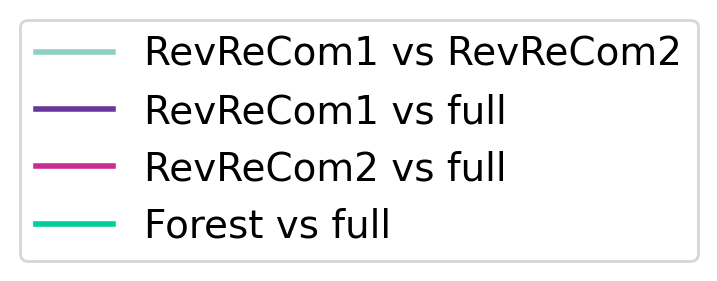}};
\node at (-4+.2,0){\includegraphics[width=3in]{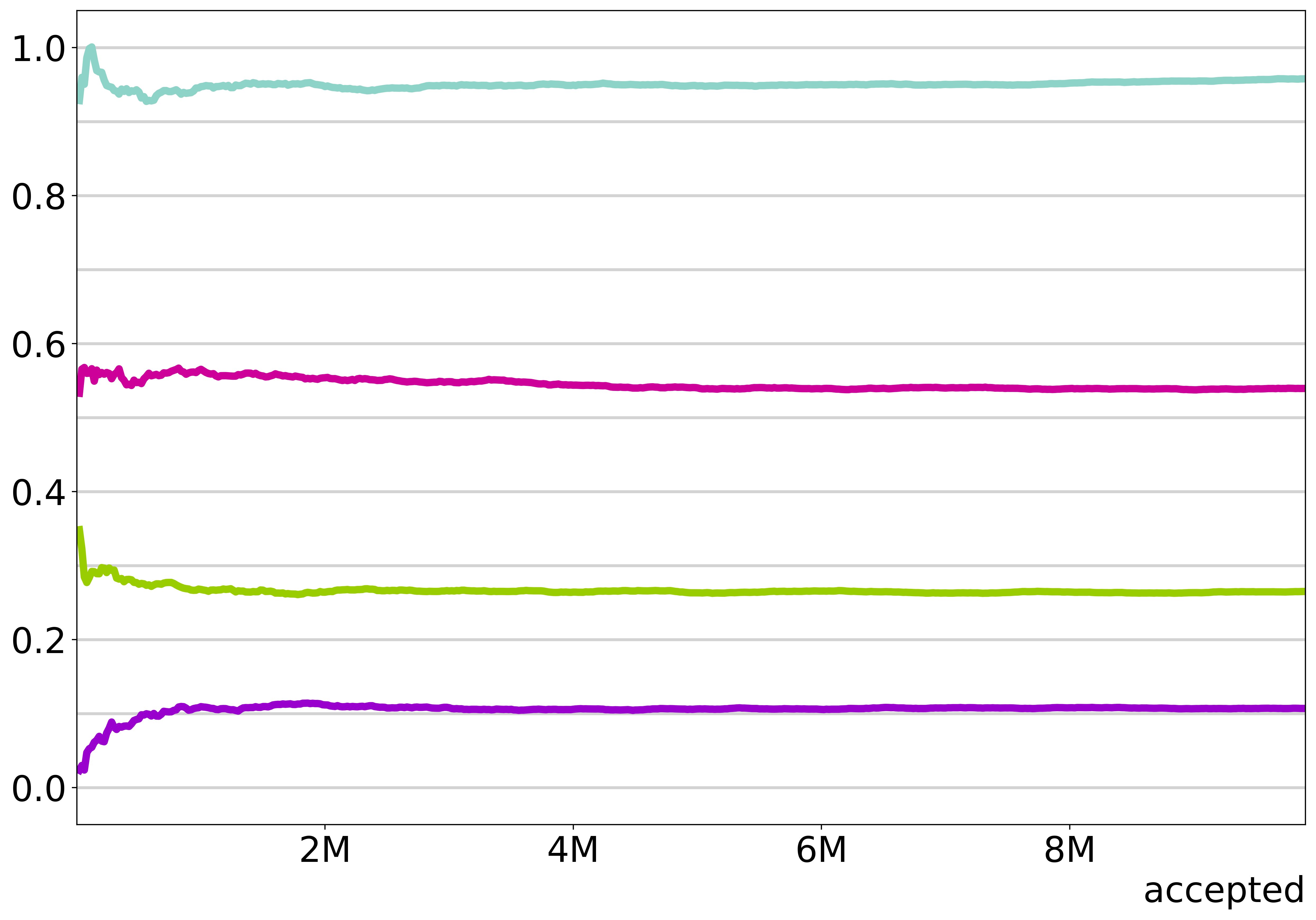}};
\node at (-1.6,1.35){\includegraphics[width=72pt]{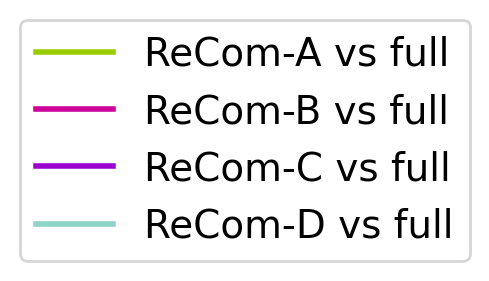}};
\node at (4,0) {\includegraphics[width=3in]{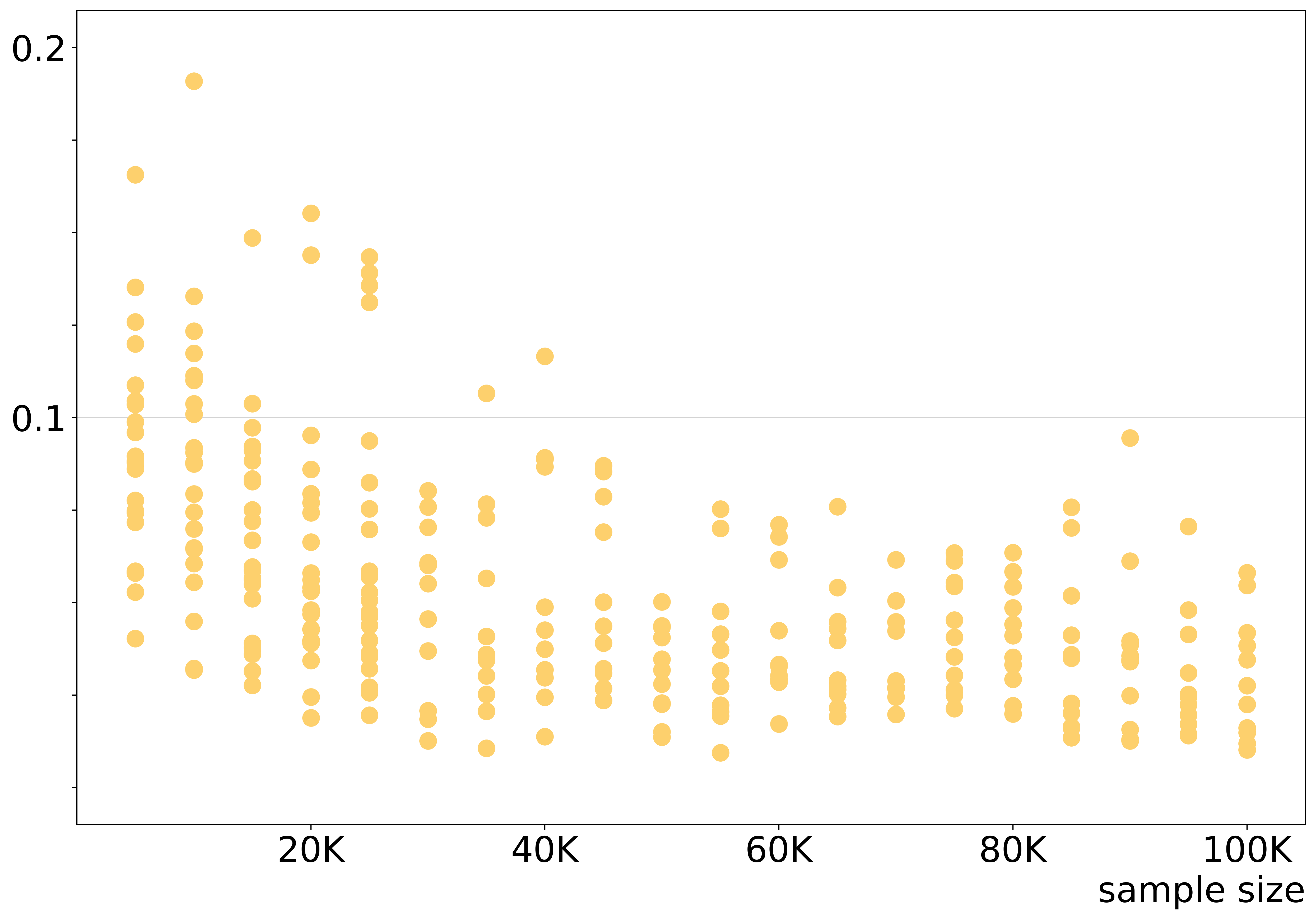}};
\node at (6.5,2){\includegraphics[width=60pt]{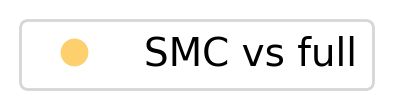}};
\end{tikzpicture}

    \caption{{\bf Convergence diagnostics for the $7\times 7$ grid.}  In this figure, we compare different methods at their largest practical sample sizes.
 TOP: A trace plot of Wasserstein distance between cut edge distributions.  Long runs are compared against each other, and against the $\pi$-weighted ground truth, showing excellent accuracy for the Markov chain methods.
 For \RevReCom, the time to 3 million accepted steps on a laptop is about ten minutes. 
 BOTTOM LEFT: The  \ReCom--A,B,C,D variants stabilize quickly, though they do not exactly converge to $\pi$.  BOTTOM RIGHT: SMC is not a Markov chain, so outputs are shown here with dots representing whole runs. Performance improves with batch size. See also Figure~\ref{fig:50x50-2}.}
	\label{fig:7x7}
\end{figure}

We will use the $7\times 7 \to 7$ districting problem as a test case:  we partition a $7\times 7$ grid into $7$ districts of $7$ units each,  which gives precisely $158,753,814$ configurations.  To date, this is the largest  districting problem that is readily manipulable in its entirety.\footnote{Imai et al. have rightly emphasized the importance of using large examples for validation \cite{ImaiEssential,SMC}, but they have tended to use much smaller test cases.  In addition, earlier demonstrations often use summary statistics unrelated to the target distribution, which have lower diagnostic value for convergence.}  
The simplest summary statistic that is closely related to the target distribution is the count of cut edges in the partition, which can vary from a minimum of 28 to a maximum of 42.  
Figure~\ref{fig:rrc-convergence} compares the distribution of cut edges (the sizes of cut-sets) in the growing ensemble to the distribution under $\pi$. 
The 100-million-step sample is visually indistinguishable from the ground-truth histogram.

Figure~\ref{fig:7x7} (top) shows all three pairwise Wasserstein distances between the ground-truth spanning tree distribution on the $7 \times 7 \rightarrow 7$ problem and two runs of \RevReCom from different seeds, using the cut edge count as the summary statistic.  The bottom left plot shows Wasserstein distance to $\pi$ for runs of \ReCom-A,B,C,D.
The SMC plots (bottom right)  use scatterplots rather than a traceplot, because SMC ensembles are built in batches rather than stepping from plan to plan over time.  But the idea is similar: each point shows the Wasserstein distance from the cut edge distribution of an SMC ensemble to the ground-truth  $\pi$-weighted distribution.

\FloatBarrier

A key takeaway from Figure~\ref{fig:7x7} is that the time it takes the two ensembles generated by \RevReCom to be similar to each other tracks with the time it takes for each to be similar to the full enumeration.  This is encouraging for the value of the multi-start heuristic when there is no full enumeration available.

\subsection{Benchmarking at full scale}\label{subsec:L1wass}

\begin{figure}
\centering
\begin{tikzpicture}
\node at (0,8.8) {\includegraphics[width=6in]{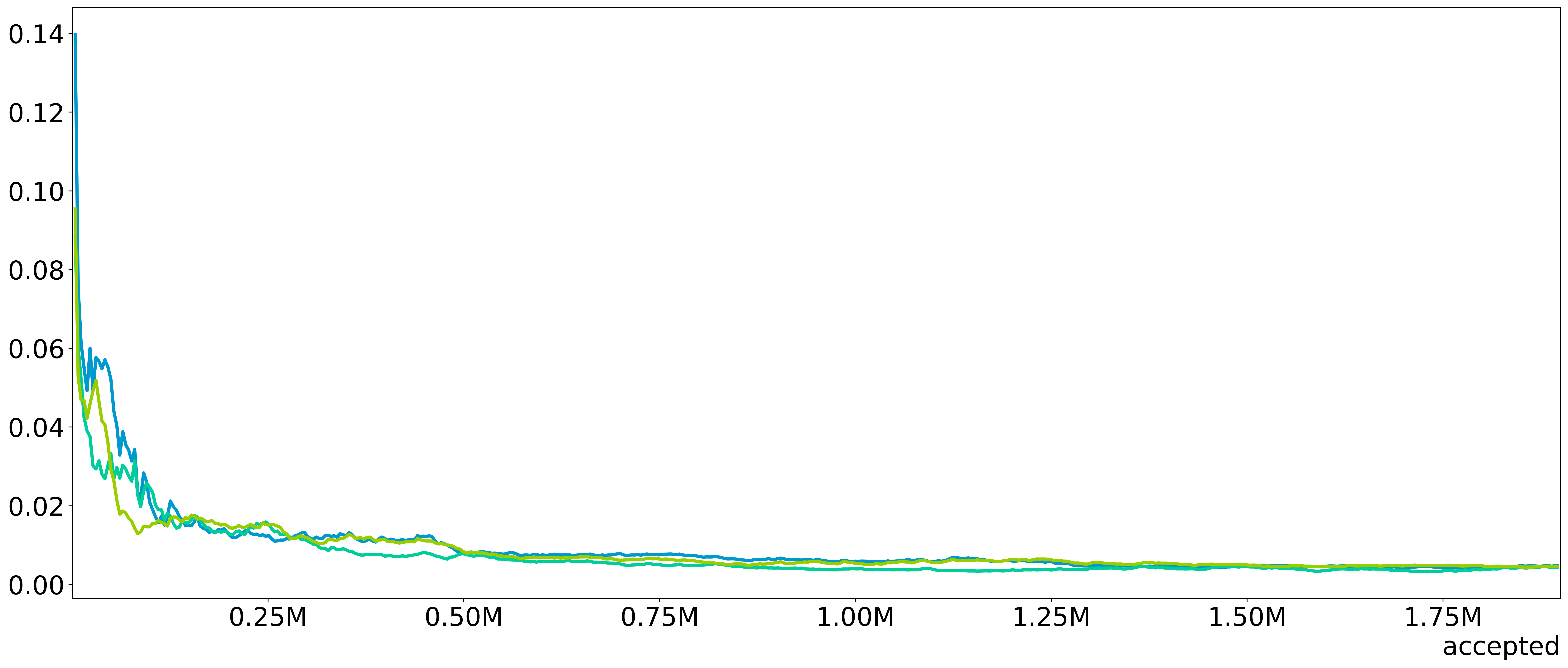}};
\node at (4.5,11) {\includegraphics[width=155pt]{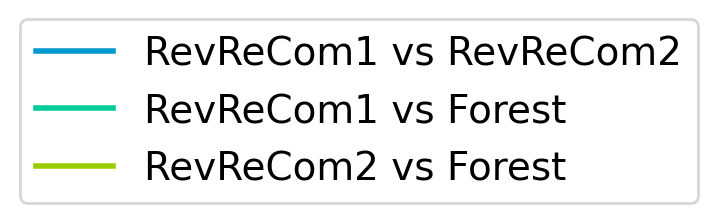}};
\node at (-7.1,12.2)  {\footnotesize Wasserstein};
\node at (0,0) {\includegraphics[width=6in]{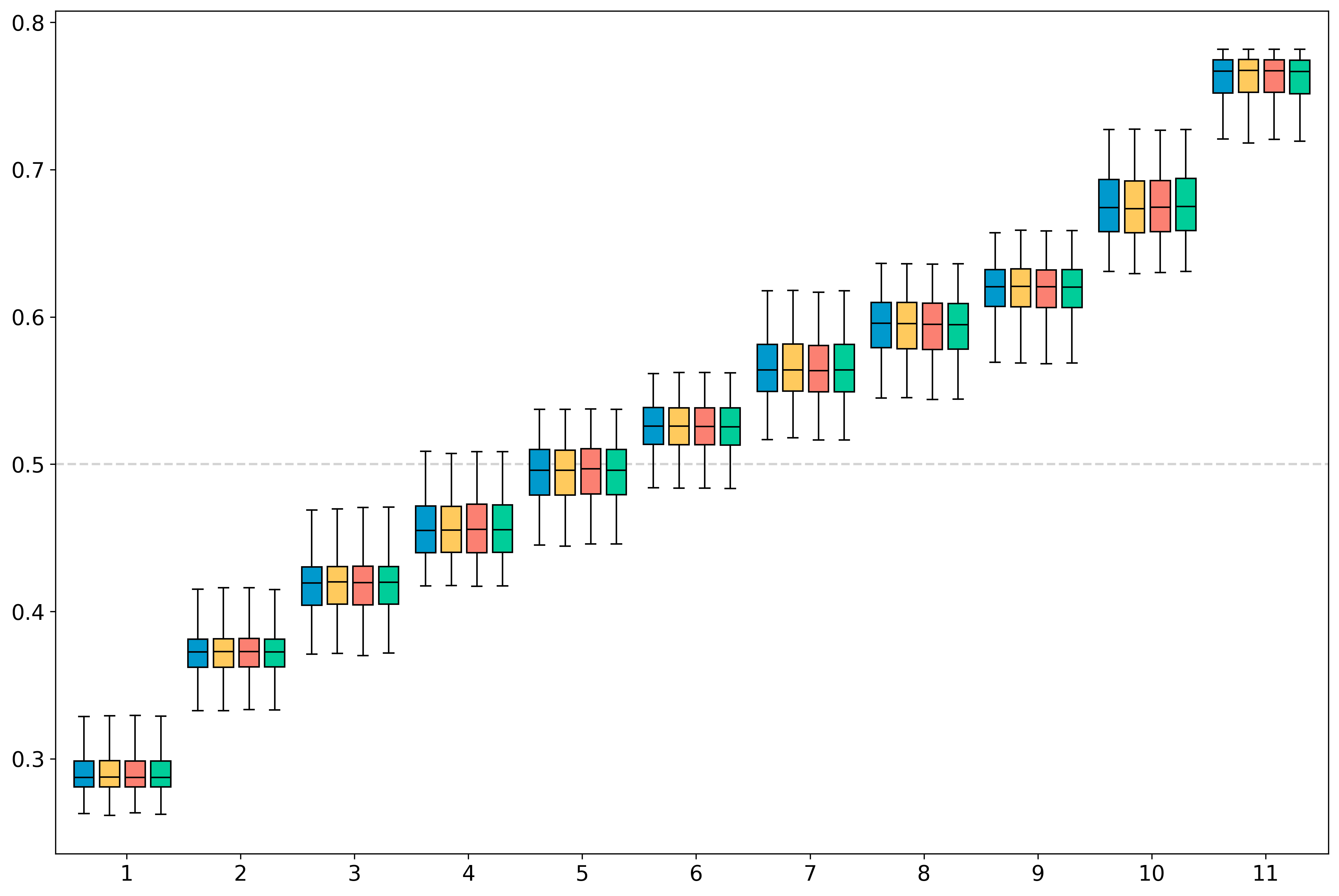}};
\node at (-5,3.7) {\includegraphics[width=90pt]{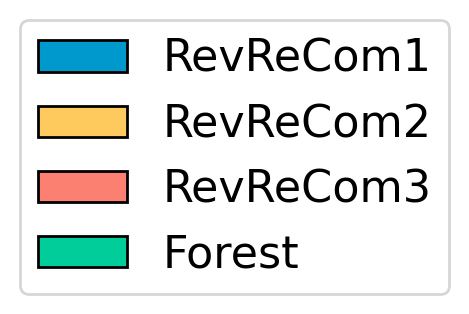}};
\node at (-7.1,5.2)  {\footnotesize Dem share};

\end{tikzpicture}

\caption{{\bf Virginia example.} TOP: The  $L^1$ Wasserstein trace plot for Democratic vote share by district (from the 2016 Presidential contest) across three different pairwise comparisons of \RevReCom and Forest \ReCom runs.
 BOTTOM: The corresponding box-and-whiskers plot showing the partisan shares by district.   
The left-most column shows the range of shares in the least Democratic district in each plan; the next column in the second-least Democratic; and so on.   In this race, Clinton received 52.8\% of the major-party share, and this plot shows that in a $\pi$-typical plan, 6/11 districts would have a Democratic advantage, 4/11 would have a Republican advantage, and the last would be very competitive, if people voted as they did in this (Clinton vs. Trump) vote pattern.}\label{fig:VA}
\end{figure}

In districting at the state level, the number of possible districting plans is so large that there is no hope of enumerating them all to get a ground-truth distribution. However, we can still compare runs across different seeds, as  in Figure~\ref{fig:VA}. 

Because our goal is to compare $k$-piece partitions, it is also natural to use statistics that attach a numerical value to each district rather than to the plan as a whole. 
To illustrate, we generate a plot in Figure~\ref{fig:VA} based on the Clinton share of the major-party presidential vote in 2016 across the districts of Virginia Congressional plans. For a given plan, we  then sort these values from lowest to highest and treat the statistic as vector-valued. 
For instance, one districting plan might give the ordered Clinton shares $(.29, .37, .41, .46, .49, .52, .55, .59, .62, .66, .75)$ across the eleven districts.
In this way we can compare plans by comparing their lowest Clinton-share district, indexed 1 across the ensemble, or the second-lowest, and so on.\footnote{It is important to remember that this means we are comparing districts that may be geographically unrelated.}
We can also visualize this in a boxplot, shown in Figure~\ref{fig:VA} (bottom) for the Virginia Congressional plans; the boxes show the 25th-75th percentile range for the value in that indexed district, the whiskers capture the 1st-99th percentile range, and the median is marked.


To get an overall convergence diagnostic, we can apply $L^1$ Wasserstein distance to these district-level values. An ensemble of plans with a district-level numerical score defines $k$ distributions over the real numbers: the first distribution is given by the values at index 1 (which are the smallest in their plans, by our order convention), the next at index 2, and so on. 
To compare two vector-valued distributions, one first computes the $k$ coordinatewise Wasserstein distances. Summing these gives the $L^1$-Wasserstein distance.  This is a  measure of the distance between two distributions, and can be computed whether the distributions come from a sample, as in an ensemble, or from a complete enumeration.
A traceplot can then be used to show the distances between time-$t$ ensembles and a static distribution, or between two time-$t$ ensembles, as in Figure~\ref{fig:VA} (top).

For Figure~\ref{fig:VA}, the starting plan is the Congressional districting plan from either 2012 or 2016, and we run the latter with two different random-number seeds. The runtime is under five hours. 
The similarity in Democratic vote statistics across the three runs lends support to the idea that 1 million accepted proposals is sufficient to produce high-quality partisan statistics by district (see also Figure~\ref{fig:50x50-2}, which suggests that even shape statistics, which are more discerning, can also be accurately obtained in tractable time).

\FloatBarrier


\subsection{Heatmaps and cross-validation}\label{sec:heatmaps}

The modifications that let \RevReCom target $\pi$ are calibrated rejection steps; not only is the acceptance rate low,  but convergence also requires many accepted steps.\footnote{On the $7\times 7$,   16.4\% of  proposals (about one in six)  were accepted in one benchmarking run. In the state-level runs,  roughly $0.05\%$ of  proposals (or one in 2000) were accepted.  See Supplemental Table~\ref{tab:rejection}.}
Thus the asymptotic guarantees come at a significant cost---not only clock time, but also handling much larger datasets---which makes this \RevReCom chain unlikely to overtake the faster \ReCom chains, or SMC runs with smaller ensemble sizes, in practical use. However, it still serves several crucial purposes: very long runs of \RevReCom give the clearest picture yet of the true spanning tree distribution, and this helps us to better evaluate the sampling distributions produced by other methods. 

\begin{figure} \centering 

\begin{tikzpicture}
\node at (-6,5.5) {\includegraphics[width=1.5in]{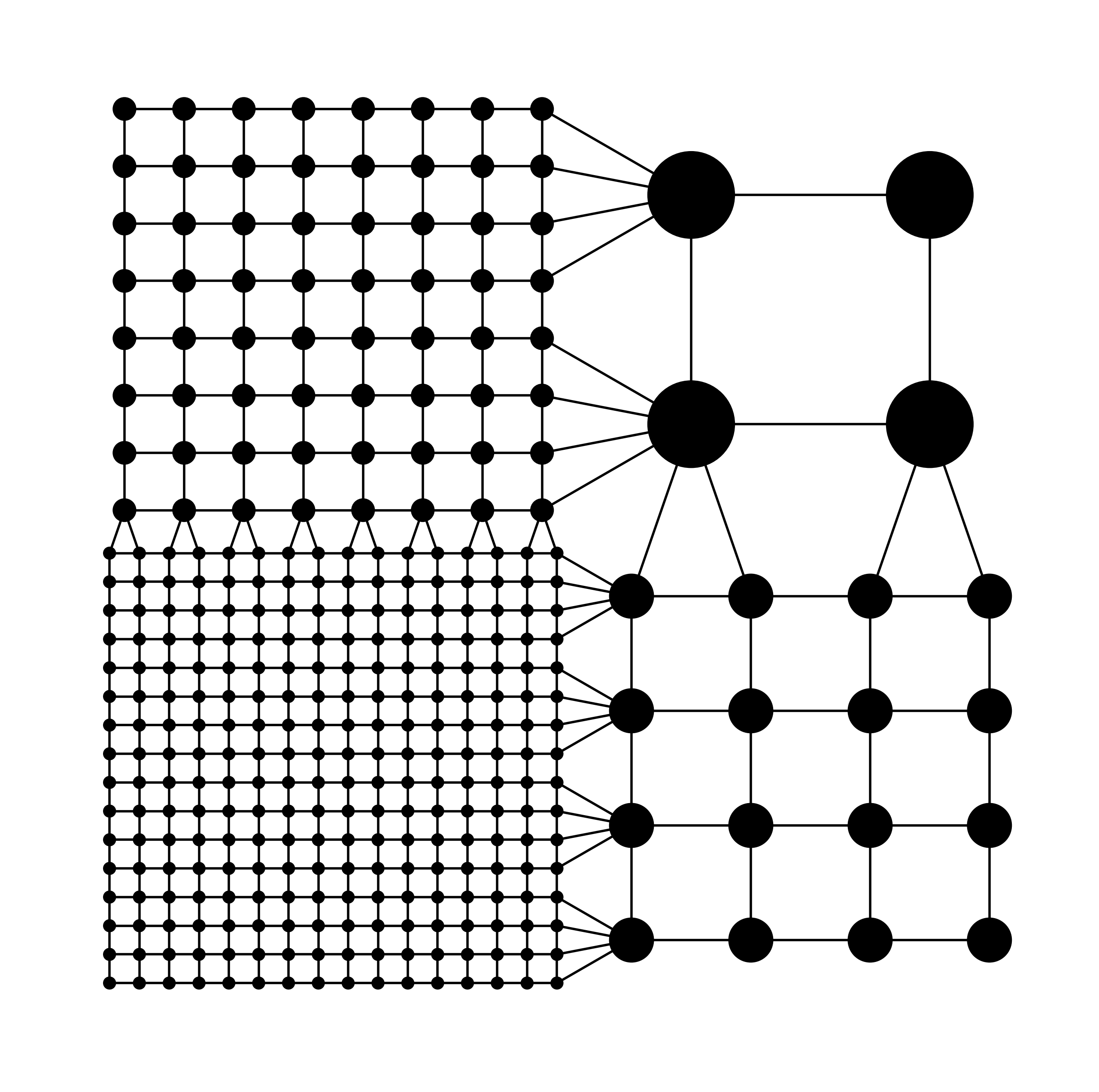}};
\node at (1,5.5) {\includegraphics[width=4in]{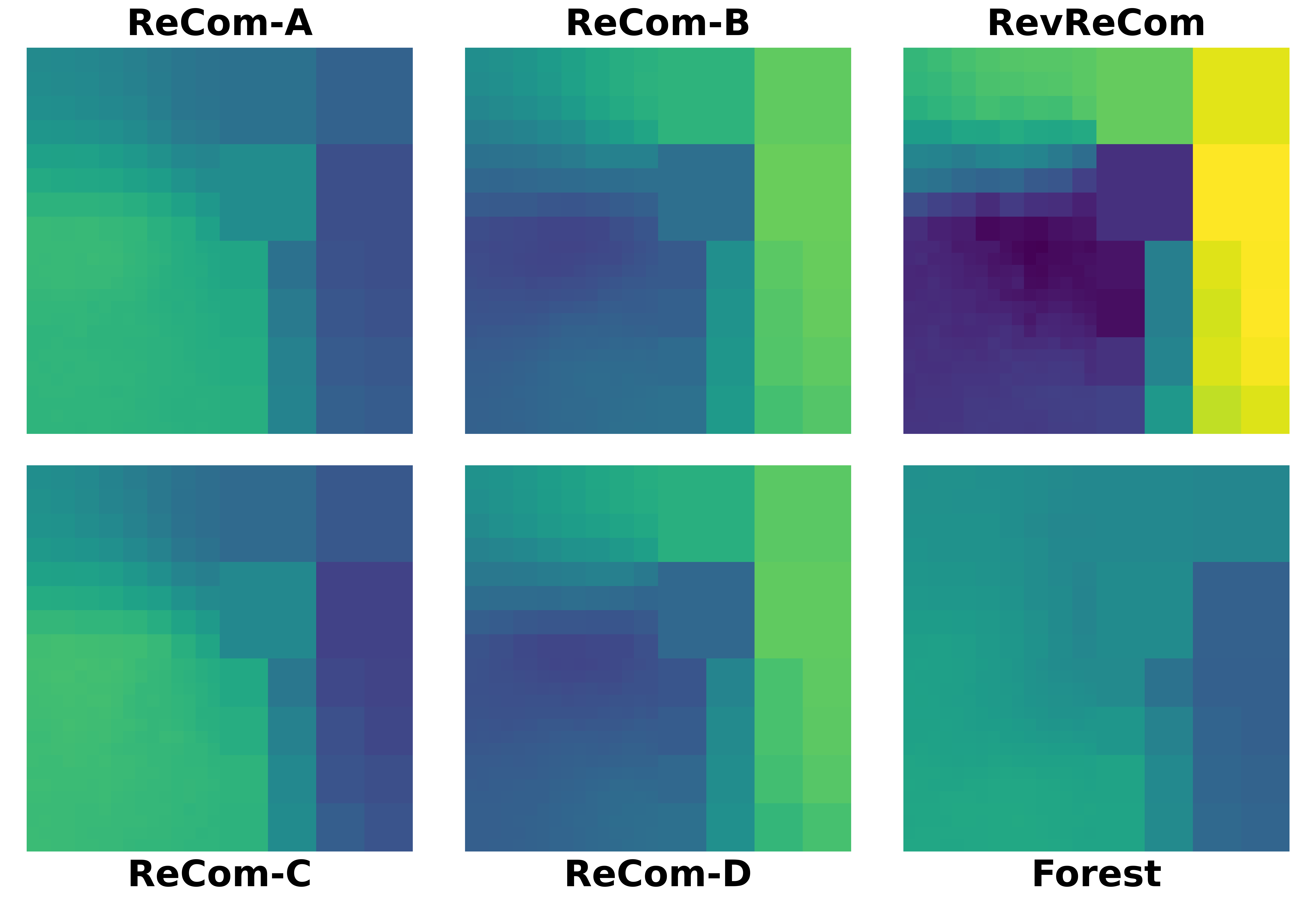}};
\node at (7,5.5) {\includegraphics[width=30pt]{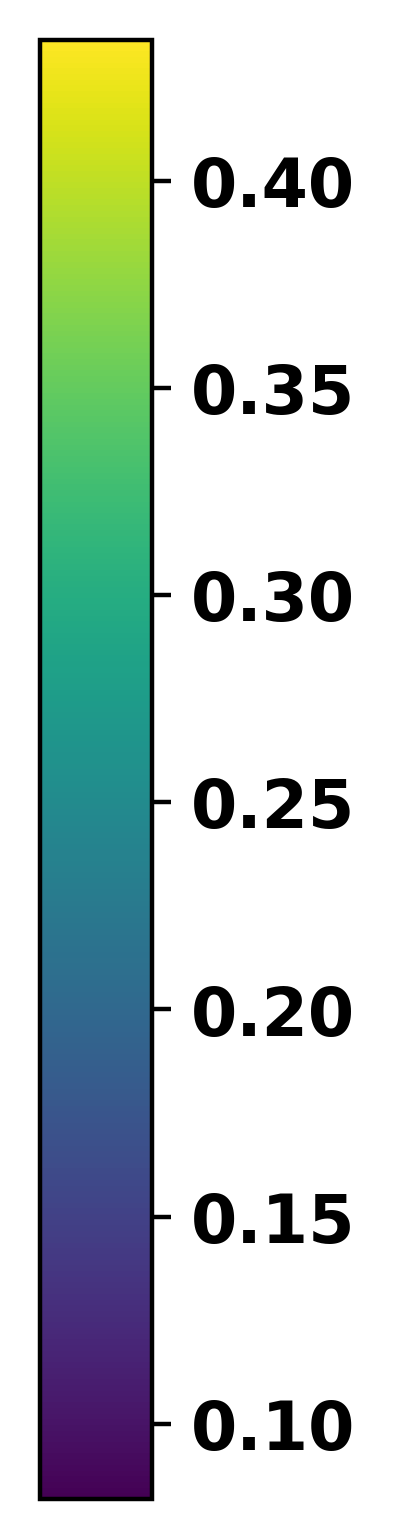}};
\node at (0,0) {\includegraphics[width=4.5in]{figSIREV/linear_multigrid_dual_graph.png}};
\node at (0,-3.6) {\includegraphics[width=6.5in]{figSIREV/linear_multigrid_heatmap_all.png}};
\node at (0,-6.7) {\includegraphics[width=4.5in]{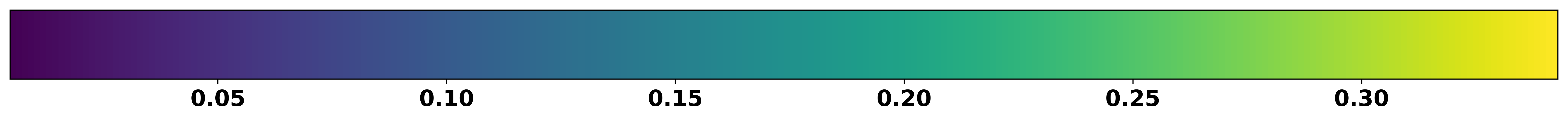}};
\node at (0,-7.7)
{\small
\begin{tabular}{cc}
\ReCom-A (cut edge, MST) & \ReCom-B (district index, MST) \\
\ReCom-C (cut edge, UST) & \ReCom-D (district index, UST)
\end{tabular}};
\end{tikzpicture}
\caption{{\bf Multiscale examples.} 
The heatmaps provide a visualization of how each random walk transits the state space of balanced partitions by showing how often each node is reassigned in the first 50,000 accepted proposals. We see that \ReCom-A and \ReCom-C, which both use cut edges to select districts to fuse, have very similar behavior, whether UST or MST is used to choose spanning trees; similarly for \ReCom-B and \ReCom-D, which both use random indices to choose districts.  
\RevReCom and Forest \ReCom flip different vertices but arrive at precisely the same steady state.}
	\label{fig:diff-scales}
\end{figure}

Comparing different methods on the same redistricting problems allows for comparisons that go beyond convergence time.
One visualization that captures some significant differences is a heatmap that shows how often nodes are reassigned to new districts in the course of a Markov chain run.
In Figure~\ref{fig:diff-scales}, two grid graphs are constructed with widely varying node weight to  exaggerate the disparities in population across nodes that can be observed in realistic districting problems.
These graphs model the partition problem on two multiscale regions that divide homogeneous rectangles into squares of different sizes.  
For the first example, each quadrant of a square has equal population, but the quadrants are subdivided into 
a $2\times 2$,  $4\times 4$, $8\times 8$, and $16\times 16$ grids, respectively.  That means that the dual graph of this tiling has 340 vertices, with the largest nodes having 64 times the population of the smallest.
This graph will be partitioned into 4 districts of equal size.
The second multiscale grid graph in this experiment has 5460 nodes, with six sections divided $2\times 2, 4\times 4,\dots, 64\times 64$, giving the largest nodes 1024 times the population of the smallest.  This graph will be partitioned into 6 districts of equal size.  (Disparities this large can be found in real geography.  In Iowa, the largest county has roughly 140 times the population of the smallest; in Texas, the ratio is over 100,000.)  We run millions of steps of each of the six Markov chains and record how often each node is reassigned.

We find that some pairs of chains (\ReCom-A and \ReCom-C, or \ReCom-B and \ReCom-D) have visually indistinguishable patterns of reassigning nodes, though those four ultimately arrive at clearly different stationary distributions. On the other hand, Forest \ReCom and \RevReCom have quite different reassignment patterns, while they converge to precisely the  same steady state.
This should raise our confidence in the statistics produced by Forest \ReCom and \RevReCom, when they agree---they are truly distinct samplers, and so their findings are not redundant but are mutually cross-validating.\footnote{This is subject to the caveat above and in \S\ref{subsec:obstruc-continued} about ergodicity; if the state space is disconnected and both Markov chain methods are initialized in the same component, then the cross-validation only applies to that component.}



\subsection{Convergence considerations}

\RevReCom, Forest \ReCom, and SMC all have (caveated) asymptotic guarantees that their sampling distributions converge to $\pi$ given large enough samples.  We briefly review the caveats that come with those guarantees and the feasibility of closely approximating $\pi$-distributed sampling in practice.
\RevReCom and SMC both use an upper bound $m$ on the number of balance edges, described in Section~\ref{subsec:approx}.
The Markov chain methods face questions of whether their state space is irreducible (i.e., connected), which is required for uniqueness of the stationary distribution.
SMC faces quickly exploding size requirements for sampling as the number of districts gets large.
All of these produce issues that might bias the distribution of a sample collected under resource constraints. 

Beyond the spanning tree distribution, all three methods can be set to target any distribution in principle, but may pay a steep price in convergence time for targeting distributions far from $\pi$.  For instance, Chatterjee and Diaconis  show in \cite{ChatDiaconis} that importance sampling methods like SMC will require an exponential increase in sample size as the target distribution deviates from the generating distribution (in KL divergence).  Similarly, \RevReCom and Forest \ReCom may experience steep increases in rejection rates when targeting other distributions, requiring longer runs to obtain the same number of accepted proposals.
The discussion of convergence obstructions continues in \S\ref{subsec:obstruc-continued}.

\FloatBarrier
\section{Discussion}

The method of ensembles allows us to measure the consequences as the rules of redistricting interact with the political geography of votes, supplying policymakers and courts with a much-needed means to study the central tendencies of districting.\footnote{Moreover, because Markov chain methods are easily amenable to heuristic optimization methods (hill-climbing, annealing, short bursts \cite{shortbursts}, and so on), we can also produce runs that seek to drive various scores and summary statistics up or down.  Local search methods give a view of the ``elasticity" of districting outcomes by probing how effective it is to guide the exploration process.  In this way we can start to understand, to the extent that we might have districting goals beyond the simple neutral consequences of the rules, how attainable those goals might be.}  
But they do not commit us to a view  that {\em neutral is fair}.
For instance, it is widely believed that some kind of urban/rural effect tends to create significant structural partisan advantages for a party with a more rural base, as we saw with Republicans in Pennsylvania. Generally, it is increasingly clear that neutrally drawn single-member plurality districts give systematic underrepresentation to racial and ethnic minority groups as they are currently geographically distributed in the United States \cite{MRL}. 
Reasonable observers can disagree about how to define fair redistricting:  is a ``blind" plan necessarily fair?  Or, to name another alternative,  policymakers might prefer plans that tend to favor representation that is proportional to popular preference---and randomized runs can help us understand if that is feasible within a given framework of rules.\footnote{See, for instance, \cite{DS-prop}.}

Spanning tree methods for sampling are appealing to courts for good reasons, including the growing scientific consensus around their construction and their increasing accessibility with lightweight, open-source computing.
Using spanning trees in the re-partition step controls the sizes of cut-sets without introducing additional parameters or weights, minimizing the need for tuning and the room for gaming basic outputs.  
Formalizing the use of the spanning tree distribution as the canonical choice for  weighting partitions lets us mutually compare methods under the fundamental constraints of population balance, contiguity, and a preference for compactness.  From there, if the specific setting calls for it, one can layer in weighting terms or other mechanisms that take local rules and priorities into account, having validated that the basic engine is effective enough to deliver on its asymptotic guarantees in practical time.  County preservation, city preservation, increasing minority groups' opportunity-to-elect, avoiding incumbent pairings,  resembling a prior map, or even respecting communities of interest collected through a public feedback process---all have been operationalized in ways that are compatible with these algorithmic methods.\footnote{That is not to say that these districting criteria have authoritative or one-size-fits-all interpretations, so it will remain important to have a canonical distribution that handles the most fundamental criteria.}
Reversible recombination, created with a small modification to the original recombination chains,  is the most powerful tool yet proposed, has formally verifiable properties, and gives particular insight into methods already in wide use in courts and in public policy.

The data demonstrations presented here confirm that the methods under study---the original recombination variants, Forest \ReCom, SMC, and \RevReCom---are all capable of producing reliable estimates for key summary statistics at the state level.  
SMC is at its best with small numbers of districts; with more than about ten districts, it can require forbiddingly large sample sizes to produce $\pi$-distributed samples (see Figure~\ref{fig:50x50-2}), and other distributions will be still harder to target.
Forest \ReCom and \RevReCom have the strongest accuracy performance (which can only be measured rigorously when ground truth is known), and produce mutually supporting estimates on both small and full-scale problems; \RevReCom scales best, passing convergence checks on the larger problems in reasonable time.

Taken together, ensemble methods give us essential tools for the 21st century as we engage in continued debate about our requirements, and our aspirations, for representative democracy.
The development of  spanning tree methods to understand redistricting is a model for scientific engagement in policy:  purpose-built tools have made clear progress on a pressing real-world problem.

\bibliographystyle{alpha}
\bibliography{sci-bib}

@article{MooreCommunity,
  author       = {Cristopher Moore},
  title        = {The Computer Science and Physics of Community Detection: Landscapes,
                  Phase Transitions, and Hardness},
  journal      = {Bull. {EATCS}},
  volume       = {121},
  year         = {2017},
  url          = {http://eatcs.org/beatcs/index.php/beatcs/article/view/480},
}

@article{DS-prop,
url = {https://doi.org/10.1515/for-2022-2064},
title = {Redistricting for Proportionality},
author = {Moon Duchin and Gabe Schoenbach},
pages = {371--393},
volume = {20},
number = {3-4},
journal = {The Forum},
doi = {doi:10.1515/for-2022-2064},
year = {2022},
}

@Incollection{GerryIntro,
  author =       "Moon Duchin",
  editor =       "Moon Duchin and Olivia Walch",
  title = "Introduction",
  booktitle =        "Political Geometry",
  subtitle =     "Rethinking Redistricting in the U.S. with Math, Law, and Everything In Between",
  year =         "2022",
  publisher =    "Birkhäuser Books",
  url =          "http://mggg.org/gerrybook",
  chapter =      "0",
  pages =        "1--28",
}

@Incollection{RoddWeigh,
  author =       "Jonathan Rodden and Thomas Weighill",
  editor =       "Moon Duchin and Olivia Walch",
  title = "Political geography and
representation: A case study of
districting in Pennsylvania",
  booktitle =        "Political Geometry",
  subtitle =     "Rethinking Redistricting in the U.S. with Math, Law, and Everything In Between",
  year =         "2022",
  publisher =    "Birkhäuser Books",
  url =          "http://mggg.org/gerrybook",
  chapter =      "5",
  pages =        "101--127",
}

@Incollection{GerryAlgs,
  author =       "Amariah Becker and Justin Solomon",
  editor =       "Moon Duchin and Olivia Walch",
  title = "Redistricting Algorithms",
  booktitle =        "Political Geometry",
  subtitle =     "Rethinking Redistricting in the U.S. with Math, Law, and Everything In Between",
  year =         "2022",
  publisher =    "Birkhäuser Books",
  url =          "http://mggg.org/gerrybook",
  chapter =      "16",
  pages =        "303--340",
}

@Incollection{Nelson,
  author =       "Garrett Dash Nelson",
  editor =       "Moon Duchin and Olivia Walch",
  title = "The Elusive Geography of
Communities",
  booktitle =        "Political Geometry",
  subtitle =     "Rethinking Redistricting in the U.S. with Math, Law, and Everything In Between",
  year =         "2022",
  publisher =    "Birkhäuser Books",
  url =          "http://mggg.org/gerrybook",
  chapter =      "11",
  pages =        "221--234",
}

@article{irred, 
	author = {Cannon, Sarah},
	title = {Irreducibility of Recombination {M}arkov Chains in the Triangular Lattice}, 
journal = {Discrete Applied Mathematics},
volume = {347},
pages = {75-130},
year = {2024}
}

@article{Tapp,
title = {Spanning Tree Bounds for Grid Graphs}, 
author = {Kris Tapp},
journal = {Electronic Journal of Combinatorics}, 
volume = {31}, 
number = {1}, 
year = {2024}
}

@unpublished{tuckerfoltz2023locked,
      title={Locked Polyomino Tilings}, 
      author={Jamie Tucker-Foltz},
      year={2023},
note = {Preprint. Available at \url{https://arxiv.org/abs/2307.15996}}
}

@unpublished{SMC-repetition,
      title={{Repetition effects in a Sequential Monte Carlo Sampler}}, 
      author={Sarah Cannon and Daryl DeFord and Moon Duchin},
      year={2024},
note = {Preprint. Available at \url{https://arxiv.org/abs/2409.19017}}
}

@unpublished{charikar2022complexity, 
title = {On the Complexity of Sampling Redistricting Plans}, 
author = {Moses Charikar and Paul Liu and Tianyu Liu and Thuy-Duong Vuong},
year = {2022},
note = {Preprint. Available at \url{https://arxiv.org/abs/2206.04883}}

}

@inproceedings{frieze2022subexponential, 
author = {Alan Frieze and Wesley Pegden},
title = {Subexponential mixing for partition chains on grid-like graphs},
booktitle = {Proceedings of the 2023 Annual ACM-SIAM Symposium on Discrete Algorithms (SODA)},
chapter = {},
pages = {3317-3329},
doi = {10.1137/1.9781611977554.ch127},
URL = {https://epubs.siam.org/doi/abs/10.1137/1.9781611977554.ch127},
year = {2023}
}

@article{Aldous,
author = {Aldous, David},
title = {On the {M}arkov chain simulation method for uniform combinatorial distributions and simulated annealing},
journal = {Probability in the Engineering and Informational Sciences},
volume = {1},
year = {1987},
pages = {33--46},
}

@article{AD, 
	author = {Aldous,  David and Diaconis, Persi},
	title = {Shuffling Cards and Stopping Times},
	journal = {The American Mathematical Monthly},
	volume = {93},
	number = {5},
	pages = {333-348},
	year = {1986},
}

@article{Brockwell,
author = {A. E Brockwell},
title = {Parallel {M}arkov chain {M}onte {C}arlo Simulation by Pre-Fetching},
journal = {Journal of Computational and Graphical Statistics},
volume = {15},
number = {1},
pages = {246-261},
year  = {2006},
publisher = {Taylor & Francis},
doi = {10.1198/106186006X100579},

URL = { 
        https://doi.org/10.1198/106186006X100579
    
},
eprint = { 
        https://doi.org/10.1198/106186006X100579
    
}

}

@Article{DS,
	author={Diaconis, Persi
	and Shahshahani, Mehrdad},
	title="Generating a random permutation with random transpositions",
	journal="Zeitschrift f{\"u}r Wahrscheinlichkeitstheorie und Verwandte Gebiete",
	year="1981",
	month="Jun",
	day="01",
	volume="57",
	number="2",
	pages="159--179",
	issn="1432-2064",
	doi="10.1007/BF00535487",
	url="https://doi.org/10.1007/BF00535487"
}

@article{BD,
	author = "Bayer, Dave and Diaconis, Persi",
	doi = "10.1214/aoap/1177005705",
	journal = "The Annals of Applied Probability",
	number = "2",
	pages = "294--313",
	publisher = "The Institute of Mathematical Statistics",
	title = "Trailing the Dovetail Shuffle to its Lair",
	url = "https://doi.org/10.1214/aoap/1177005705",
	volume = "2",
	year = "1992"
}

@book{diaconis-book,
	author = {Diaconis, Persi},
	title = {Group Representations in Probability and Statistics},
	year = {1988},
publisher = {Institute of Mathematical Statistics Lecture Notes},
	series = {Monograph Series},
	volume = {11},
}

@article{Doss,
	author = "Doss, Charles R. and Flegal, James M. and Jones, Galin L. and Neath, Ronald C.",
	doi = "10.1214/14-EJS957",
	journal = "Electronic Journal of Statistics",
	number = "2",
	pages = "2448--2478",
	publisher = "The Institute of Mathematical Statistics and the Bernoulli Society",
	title = "Markov chain {M}onte {C}arlo estimation of quantiles",
	url = "https://doi.org/10.1214/14-EJS957",
	volume = "8",
	year = "2014"
}

@article{KW, 
title = { What is a complex graph?}, 
author = {J. Kim and T. Wilhelm},
journal = {Physics {A}}, 
volume = {387},
year = {2008},
pages = {2637--2652},
}

@article{Lyons, 
author = {Russell Lyons},
journal = {Combinatorics, Probability, and Computing}, 
title = {Asymptotic enumeration of spanning trees}, 
volume = {14},
number = {4}, 
year = {2005}, 
pages = {491--522},
}

@article{DukeReCom1, 
	author = {Eric Autry and Daniel Carter and Gregory Herschlag and Zach Hunter and Jonathan C. Mattingly},
	title =  {Metropolized Forest Recombination for {M}onte {C}arlo Sampling of Graph Partitions},
	journal = {SIAM Journal on Applied Mathematics},
	volume = {83},
	number = {4},
	pages = {1366-1391},
	year = {2023},
	doi = {10.1137/21M1418010}
}

@article{DukeReCom2, 
	author = {Eric A. Autry and Daniel Carter and Gregory Herschlag and Zach Hunter and Jonathan C. Mattingly},
	title = {Metropolized Multiscale Forest Recombination for Redistricting}, 
	journal = {Multiscale Modeling \& Simulation},
	volume = {19},
	number = {4},
	pages = {1885-1914},
	year = {2021},
	URL = {https://doi.org/10.1137/21M1406854}
}

@article{SMC,
author = {Cory McCartan and Kosuke Imai},
title = {{Sequential Monte Carlo for sampling balanced and compact redistricting plans}},
volume = {17},
journal = {The Annals of Applied Statistics},
number = {4},
publisher = {Institute of Mathematical Statistics},
pages = {3300 -- 3323},
year = {2023},
doi = {10.1214/23-AOAS1763},
URL = {https://doi.org/10.1214/23-AOAS1763}
}

@article{ImaiEssential, 
author = {Benjamin Fifield and Kosuke Imai and Jun Kawahara and Christopher T. Kenny}, 
year = {2020}, 
title = {The Essential Role of Empirical Validation in Legislative Redistricting Simulation},
journal = {Statistics and Public Policy},
volume = {7}, 
number = {1}, 
pages = {52--68}
}

@article{Gelman1992Inference, 
 ISSN = {08834237},
 URL = {http://www.jstor.org/stable/2246093},
 author = {Andrew Gelman and Donald B. Rubin},
 journal = {Statistical Science},
 number = {4},
 pages = {457--472},
 publisher = {Institute of Mathematical Statistics},
 title = {Inference from Iterative Simulation Using Multiple Sequences},
 urldate = {2023-08-28},
 volume = {7},
 year = {1992}
}

@article{SciData-alarm,
	author = {McCartan, Cory and Kenny, Christopher T. and Simko, Tyler and Garcia, George and Wang, Kevin and Wu, Melissa and Kuriwaki, Shiro and Imai, Kosuke},
	date = {2022/11/11},
	doi = {10.1038/s41597-022-01808-2},
	id = {McCartan2022},
	isbn = {2052-4463},
	journal = {Scientific Data},
	number = {1},
	pages = {689},
	title = {Simulated redistricting plans for the analysis and evaluation of redistricting in the United States},
	url = {https://doi.org/10.1038/s41597-022-01808-2},
	volume = {9},
	year = {2022}}

@article{Vehtari2021Rank, 
author = {Aki Vehtari and Andrew Gelman and Daniel Simpson and Bob Carpenter and Paul-Christian Bürkner}, 
title = {Rank-Normalization, Folding, and Localization: An Improved $\hat{R}$ for Assessing Convergence of {MCMC} (with Discussion)}, 
journal = {Bayesian Analysis}, 
volume = {16}, 
number = {2}, 
paes = {667 - 718},
year = {June 2021},
url = {https://doi.org/10.1214/20-BA1221}
}

@article{ChatDiaconis,
author = {Sourav Chatterjee and Persi Diaconis},
title = {The sample size required in importance sampling}, 
journal = {Annals of Applied Probability},
volume = {28},
number = {2},
year = {2018}, 
pages = {1099--1135},
}

@article{MCMCRev, 
author = {Persi Diaconis},
title = {The {M}arkov chain {M}onte {C}arlo revolution}, 
journal = {Bulletin of the American Mathematical Society},
volume = {46},
year = {2009}, 
pages = {179--205},
}

@article{DuchinTenner,
title = {Discrete geometry for electoral geography},
journal = {Political Geography},
volume = {109},
pages = {103040},
year = {2024},
issn = {0962-6298},
doi = {https://doi.org/10.1016/j.polgeo.2023.103040},
url = {https://www.sciencedirect.com/science/article/pii/S0962629823002184},
author = {Moon Duchin and Bridget Eileen Tenner}
}

@article{shortbursts, 
	author = {Sarah Cannon and Ari Goldbloom-Helzner and Varun Gupta and JN Matthews and Bhushan Suwal}, 
	title = {Voting Rights, {M}arkov Chains, and Optimization by Short Bursts}, 
	journal = {Methodology and Computing in Applied Probability}, 
	year = {2023}, 
	volume = {25}, 
	number = {1}
}

@unpublished{NDS1,
title = {Complexity and Geometry of Sampling Connected Graph Partitions}, 
author = {Elle Najt and Daryl De{F}ord and Justin Solomon},
note = {Available at \url{https://arxiv.org/abs/1908.08881}}
}

@article{NDS2,
  title = {Empirical sampling of connected graph partitions for redistricting},
  author = {Najt, Elle and DeFord, Daryl and Solomon, Justin},
  journal = {Phys. Rev. E},
  volume = {104},
  issue = {6},
  pages = {064130},
  numpages = {17},
  year = {2021},
  month = {Dec},
  publisher = {American Physical Society},
  doi = {10.1103/PhysRevE.104.064130},
  url = {https://link.aps.org/doi/10.1103/PhysRevE.104.064130}
}

@article{KarpLubyMadras,
	title = {Monte-{C}arlo approximation algorithms for enumeration problems},
	journal = {Journal of Algorithms},
	volume = {10},
	number = {3},
	pages = {429-448},
	year = {1989},
	author = {Richard M Karp and Michael Luby and Neal Madras},
}

@Unpublished{clelland2021compactness,
	title={Compactness statistics for spanning tree recombination}, 
	author={Jeanne N. Clelland and Nicholas Bossenbroek and Thomas Heckmaster and Adam Nelson and Peter Rock and Jade VanAusdall},
	year={2021},
	note={Preprint. Available at \url{https://arxiv.org/abs/2103.02699}}
}

@article{procacciaTuckerFoltz2021compactness, 
	title = {Compact Redistricting Plans Have Many Spanning Trees}, 
	author = {Ariel D. Procaccia and Jamie Tucker-Foltz},
    journal = {ACM-SIAM Symposium on Discrete Algorithms (SODA)}, 
    year = {2022}
}

@misc{AF, 
AUTHOR = {Aldous, David and Fill, James Allen},
TITLE = {Reversible {M}arkov Chains and Random Walks on Graphs},
YEAR = {2002},
NOTE = {Unfinished monograph, recompiled 2014, available 
at \url{http://www.stat.berkeley.edu/~aldous/RWG/book.html}}
}

@book{feller,
	author = {Feller, William},
	title = {An Introduction to Probability Theory and Its Applications},
	publisher = {Wiley},
	volume = {1},
	year = {1968}
}

@article{ReCom, 
author = {DeFord, Daryl and Duchin, Moon and Solomon, Justin},
	journal = {Harvard Data Science Review},
	number = {1},
	year = {2021},
	publisher = {},
	title = {Recombination: A {Family} of {Markov} {Chains} for {Redistricting}},
	volume = {3},
}

@article{POM,
author = {Mason A. Porter and Jukka-Pekka Onnela and Peter J. Mucha},
title = {Communities in Networks}, 
journal = {Notices of the American Mathematical Society}, 
volume = {56}, 
number = {9}, 
year = {2009}, 
pages = {1082--1097, 1164--1166}
}

@article{Geyer, 
author = {Charles J. Geyer},
title = {{Practical Markov Chain Monte Carlo}},
volume = {7},
journal = {Statistical Science},
number = {4},
pages = {473--483},
year = {1992}
}

@inproceedings{wilson,
author = {Wilson, David Bruce},
title = {Generating Random Spanning Trees More Quickly than the Cover Time},
year = {1996},
url = {https://doi.org/10.1145/237814.237880},
doi = {10.1145/237814.237880},
booktitle = {Proceedings of the Twenty-Eighth Annual ACM Symposium on Theory of Computing},
pages = {296–303},
numpages = {8},
series = {STOC}
}

@article{akitaya2022reconfiguration_recombination,
title = {Reconfiguration of connected graph partitions via recombination},
journal = {Theoretical Computer Science},
volume = {923},
pages = {13-26},
year = {2022},
author = {Hugo A. Akitaya and Matias Korman and Oliver Korten and Diane L. Souvaine and Csaba D. Tóth}
}

@unpublished{Gerrychain, 
	author = {{MGGG Redistricting Lab}},
	title = {GerryChain}, 
	note = {Python Library. \url{https://github.com/mggg/GerryChain}}
}

@unpublished{GerryChainRust,
  author       = {Parker Rule},
  title        = {FRCW},
  note = {Rust Implementation of Reversible ReCom. \url{https://github.com/mggg/frcw.rs}}
}

@unpublished{replication,
  author       = {Peter Rock and Parker Rule},
  title        = {Reversible ReCom Replication},
  note = {GitHub Repository. \url{https://github.com/mggg/RRC-Replication}}
}

@unpublished{Forest-repos, 
	author = {Gregory Herschlag},
	title = {mergeSplitCodeBase}, 
	note = {Python Library and Julia Library. \url{https://git.math.duke.edu/gitlab/gjh/mergesplitcodebase}  \url{https://git.math.duke.edu/gitlab/quantifyinggerrymandering/multiscalemapsampler-public}}
}

@unpublished{Redist-repo, 
	author = {Christopher Kenny and Cory McCartan and Ben Fifield and Kosuke Imai},
	title = {redist: Simulation Methods for Legislative Redistricting}, 
	note = {GitHub Repository. \url{https://github.com/alarm-redist/redist/}}
}

@article{MRL, 
author = {Duchin, Moon and Spencer, Douglas M.}, 
title = {Models, {R}ace, and the {L}aw}, 
journal = {The Yale Law Journal}, 
volume = {130}, 
pages = {744-797},
year = {2021}
}

@inproceedings{splittable,
author = {Cannon, Sarah and Pegden, Wesley and Tucker-Foltz, Jamie},
title = {Sampling Balanced Forests of Grids in Polynomial Time},
year = {2024},
isbn = {9798400703836},
publisher = {Association for Computing Machinery},
address = {New York, NY, USA},
url = {https://doi.org/10.1145/3618260.3649699},
doi = {10.1145/3618260.3649699},
booktitle = {Proceedings of the 56th Annual ACM Symposium on Theory of Computing},
pages = {1676–1687},
numpages = {12},
location = {, Vancouver, BC, Canada, },
series = {STOC 2024}
}

\section{Acknowledgements}

The authors are grateful to the developer team behind GerryChain, with special thanks to Daryl DeFord, Max Hully, JN Matthews, Bhushan Suwal, Anthony Pizzimenti, Max Fan, and Peter Rock, as well as other members and collaborators of the Data and Democracy Lab.  We thank Kosuke Imai for suggesting that we include sequential Monte Carlo (SMC) comparisons alongside Markov chain methods, and also thank Cory McCartan for his help making sure we were using SMC as intended.

{\bf Funding}. The authors are deeply grateful for funding support from multiple sources. S.C. acknowledges NSF grants DMS-1803325 and CCF-2104795;
M.D. acknowledges NSF DMS-2005512;
D.R. acknowledges NSF grants CCF-1733812 and CCF-2106687.  S.C., M.D., and D.R. are grateful to the Simons-Laufer Mathematical Sciences Institute, where we were research members together while this manuscript was in preparation.

{\bf Author Contributions}.
All authors were involved in conceptualization, methodology, investigation, and writing.  M.D. and P.R. were engaged in data curation, and P.R. is the principal software developer for this project.  Author order is alphabetical, following the convention in mathematics.

{\bf Competing Interests}. The authors have no competing interests. 
M.D.~served as an expert in {\em League of Women Voters v. Pennsylvania} (2018), the case that provides context for Figure~\ref{fig:PA-compare}, as well as in {\em Allen v. Milligan} (2023) and numerous other cases in the current redistricting cycle.

{\bf Data and Materials Availability}.  

Code for these experiments is publicly available at \href{https://github.com/mggg/RRC-Replication}{\tt https://github.com/mggg/RRC-Replication} and \href{https://github.com/mggg/frcw.rs}{\tt github.com/mggg/frcw.rs}. 


\newpage
\appendix

\section{Legal reception}\label{app:scotus}
By looking at just two recent cases---{\em Rucho v. Common Cause}, a 2019 partisan gerrymandering case from North Carolina, and {\em Allen v. Milligan}, a 2023 racial vote dilution case from Alabama---we can find all nine current U.S. Supreme Court justices giving weight to redistricting algorithms as helpful evidence, and specifically citing work that uses Markov chain methods. 
The selections quoted here show an emphasis on finding examples or a benchmark, not a putatively best or optimized choice.  In addition, courts have looked for reassurance that this baseline for comparison does not depend sensitively on idiosyncrasies of the method.

\paragraph{Kagan in {\em Rucho}  dissent} (joined by Sotomayor, Ginsburg, and Breyer)
\begin{itemize}
    \item ``The approach... begins by using advanced computing technology to randomly generate a large collection of districting plans that incorporate the State’s physical and political geography and meet its declared districting criteria, except for partisan gain. ... The further out on the tail, the more extreme the partisan distortion and the more significant the vote dilution."  (p18-19)
    \item ``The point is that the assemblage of maps, reflecting the characteristics and judgments of the State itself, creates a neutral baseline from which to assess whether partisanship has run amok." (p23-24)
\end{itemize}

\paragraph{Roberts in {\em Milligan}  majority opinion} 
(joined by Kagan, Sotomayor, Jackson, and Kavanaugh)
\begin{itemize}
    \item Evidence toward the Gingles 1 precondition: 
 ``the `randomized algorithms' [plaintiffs' expert] employed `found plans with two majority-black districts in literally thousands of different ways.'" (footnote 7, p26)
  \item Kavanaugh's concurring opinion asserts that ``computer simulations might help detect the presence or absence of intentional discrimination" (p3).
 \item The Roberts opinion is  open to the use of ensembles in voting rights cases (p28), while  opting not to {\em require} them for claims under \S~2 of the Voting Rights Act, which is concerned with minority groups' opportunity to elect candidates of choice.
\end{itemize}

\paragraph{Thomas in {\em Milligan} dissent} (joined by Barrett, Gorsuch, and Alito)
\begin{itemize}
\item ``In arguing that a vote-dilution claim requires judging a State’s plan relative to an undiluted benchmark to be drawn from the totality of circumstances---including, where probative, the results of districting simulations---the State argues little more than what we have long acknowledged."  (footnote 15, p31)
\item Thomas calls \ReCom-based ensemble evidence from \cite{MRL} ``surely probative" (p23).
\item Alito also writes in a separate opinion that ensembles provide ``strong evidence" for the Alabama case (p13).  
\item In fact, the dissenting bloc writes in favor of {\em requiring} this kind of ensemble evidence for a Voting Rights Act claim.
\end{itemize}

\section{Theory}\label{app:theory}

\subsection{Potential obstructions to convergence}\label{subsec:obstruc-continued}

\subsubsection{Reducibility}

From the detailed balance expressions given in \S\ref{subsec:exact} for the perfect balance case and in \S\ref{subsec:approx} for the approximate balance case, we know that the spanning tree distribution $\pi$ is a steady state of \RevReCom.  However, the fundamental theorem that guarantees that this is the unique stationary distribution, and that all other distributions converge to $\pi$ under iterations of the Markov process, requires the hypothesis that the system is {\em ergodic}, i.e., aperiodic (the gcd of all lengths of closed cycles is 1) and irreducible (it is possible to transition from any state to any other state in finite forward time). In all Markov chains used to sample graph partitions on real-world data, including all the ones described here,  aperiodicity is immediate because there are large cliques, but proofs of irreducibility have remained elusive.   
We note that \ReCom  has the same state space as \RevReCom, with the same allowed transitions but different transition probabilities, so irreducibilty for \ReCom and \RevReCom are equivalent. Recombination moves include flips as a special case, so the state spaces for flip chains are strictly more likely to be reducible, i.e., disconnected.

There is mounting evidence that recombination chains are irreducible under realistic sampling conditions on real-world dual graphs, despite the difficulties in proving it.  
The main challenge is the population constraints: the tighter the allowed deviation, the more likely it is that the valid moves no longer connect the state space. As population constraints loosen, irreducibility is known: Akitaya et al. have shown that the chain is irreducible when the dual graph is Hamiltonian and districts can shrink arbitrarily small and grow up to double their ideal size~\cite{akitaya2022reconfiguration_recombination}. This is because allowed moves can create some large districts and several single-vertex districts, and moving through such configurations allows greater flexibility. A recent result of Cannon proves irreducibility under much tighter balance conditions ($\pm 1$ from ideal size), but is limited to the special case of $k=3$ districts on a triangular lattice~\cite{irred}. There are also positive irreducibility results by Charikar et al. under tight size constraints for specific, highly structured~examples~\cite{charikar2022complexity}.
Many of the negative results are for unrealistic, stylized graph ``gadgets," such as by Akitaya et al. for planar graphs with many more large faces than one would expect in a typical geography dual graph~\cite{akitaya2022reconfiguration_recombination}. 

Small rectangular grids produce many interesting examples when there are few units per district.
Tucker-Foltz finds a range of interesting behavior in small finite grid examples where districts have 3-5 units each~\cite{tuckerfoltz2023locked}. 
He gives several examples of tilings of finite regions by $m$-ominoes, for $m=3,4,5$, that are {\em recombination rigid}, meaning that they are not connected by any perfect-balance recombination transition to any other configuration, so they are isolated points in the state space; he also shows that certain grid examples, like the $6\times 6\to 12$ problem, have state spaces with multiple large connected components.
The $6\times 6\to 3$ problem (a $6\times 6$ grid divided into $k=3$ exactly balanced districts of 12 units each) is known to be ergodic for recombination, but every known proof involves extensive case analysis. For the 
$6 \times 6\to 6$ and 
$7\times 7 \to 7$ problems, all known proofs of irreducibility use brute-force computation. This direct verification of irreducibility is computationally infeasible for the larger dual graphs arising from real-world instances.

Realistic geography dual graphs are usually planar (though exceptions can arise when the pieces themselves are disconnected) and tend to look like near-triangulations of simply connected domains, with hundreds or thousands of units per district.
No examples of disconnected state spaces  are known when the dual graph has small perimeter compared to interior (as in grids and lattices and all known real-world examples) and the number of units per district is at least 6.  
Thus, recombination is widely believed to be irreducible on geography dual graphs under legally reasonable values of the population deviation tolerance $\epsilon$, such as $\epsilon=.01$ for Congressional districts and $\epsilon=.05$ for smaller legislative districts.

Without a proof of irreducibility, we only know that a \RevReCom chain initialized at a particular state converges to the spanning tree distribution on its  connected component of the state space.  This provides another reason for verifying that samples seeded at multiple different starting points have similar properties---also known as the multistart heuristic, as in Figure~\ref{fig:VA}.  
We note that the problem of finding starting points, or seeds, can be a significant challenge facing MCMC practitioners; one possible application of SMC, even long before its samples approximate any target distribution, is to generate starting points for Markov chain exploration.


\subsubsection{Ancestor extinction in SMC}
Due to its iterative structure of marking districts, the SMC method from \cite{SMC} faces certain convergence obstructions that are not faced by the Markov chain methods.  Each of the $k$ districts is selected in a separate ``generation" of draws from partial partitions, creating the very likely prospect of so-called {\em ancestor extinction}:  multiple plans in the $i$th generation will likely be drawn from the same parent in generation $i-1$, and the concentration compounds generationally, so that a set of identical districts are likely to be found in common in a high share of plans finally produced, while other districts drawn in early generations are never found in the final output.  For example, a recent expert report in New York legislative redistricting used SMC to generate samples of state Senate plans with $k=63$ districts.  A rebuttal report replicated the SMC ensemble used by the plaintiffs' expert and found that 31 districts out of 63 were repeated {\em exactly identically} in most of the plans (3129 out of 5000) (Affidavit of Kristopher Tapp in {\em Harkenrider v. Hochul}, see \href{https://www.democracydocket.com/wp-content/uploads/2022/02/06-2022-04-04-AFFIRMATION-OF-ALEXANDER-GOLDENBERG-IN-SUPPORT-OF-APPELLANTS-MOTION-BY-ORDER-TO-SHOW-CAUSE.pdf}{brief}). 

The SMC authors are aware of these issues and take mitigating steps in their own work. One such such step is to {\em modularize} large states into smaller pieces in an arbitrary way, though this may change the sampling distribution overall.  Another step is to combine subsamples from separate batches rather than reporting single batches.  Both of these strategies are employed  in the \href{https://github.com/alarm-redist/fifty-states/blob/main/analyses/TX_cd_2020/doc_TX_cd_2020.md}{Texas}, \href{https://github.com/alarm-redist/fifty-states/blob/main/analyses/FL_cd_2020/doc_FL_cd_2020.md}{Florida}, and \href{https://github.com/alarm-redist/fifty-states/blob/main/analyses/CA_cd_2020/doc_CA_cd_2020.md}{California} ensembles from their 50-state project, now published in {\em Scientific Data} \cite{SciData-alarm}.  These techniques are decoupled from theoretical guarantees about the sampling distribution, so we have opted to conduct the largest SMC runs possible (see ``largest practical sample sizes," \S\ref{sec:samplers}) and report the resulting ensembles in full.

In addition, SMC uses a correction to account for the fact that it samples labeled rather than unlabeled partitions; 
this essentially requires a second round of importance sampling to estimate a corrective factor for the first, controlled by an additional parameter called {\tt est\_label\_mult}.  There is an ``exploration" parameter {\tt seq\_alpha} and various additional parameters to loosen constraints if the process is getting stuck ({\tt pop\_temper}, {\tt final\_infl}).
All of these can help produce diverse-looking samples in a reasonable time, but will likely significantly expand the sample size needed to get batches that  resemble draws from $\pi$.
As a result, we would expect to need batch sizes of many millions in order to get approximately $\pi$-distributed samples on state-sized problems, especially with large numbers of districts---currently out of reach of laptop computing.
These issues are explored further in \cite{SMC-repetition}.


\subsection{Constructing Markov chain ensembles}\label{subsec:burnin}
\label{sec:burnin_subsample_samplesize}


MCMC is frequently used to collect a sample of configurations; as noted in the main text,  ensembles collected from MCMC runs will be used to study proposed plans comparatively. Recall from \S\ref{sec:goals} that our ensembles will be built by collecting each plan visited by the random walk. We now provide additional details and context for that choice. 

\paragraph{Continuous observation.}
In many  applications, researchers will employ parameters $s_1$ and $s_2$ to implement {\em burn-in} and {\em sub-sampling}:  a Markov chain process will skip the first $s_1$ states before adding a state to the ensemble; subsequently, every $s_2^{\rm th}$ state visited by the chain will be added.  
In principle, when $s_1$ is set to the mixing time of the Markov process and $s_2$ is set as its relaxation time, this produces (approximately)  uncorrelated samples from the stationary distribution. 
(Here  the {\it relaxation time} $t_{rel}$  is the inverse of the spectral gap of the transition matrix; it is closely related to the mixing time $t_{mix}$ at which the sample approaches its steady state.)
For real applications that lack rigorous bounds on mixing time, some authors have argued in favor of {\em continuous observation} that records every step encountered by the chain.  
For example, Aldous   proves in \cite{Aldous} that for estimating means, continuous observation for $2 k t_{rel}$ steps after a burn-in phase is at least as good as taking $k$ samples, each at time $s_2 = 2 t_{rel}$ apart.   Similarly, Geyer argues in \cite{Geyer} that in practice, one long enough run of the Markov chain with continuous observation suffices for estimating  quantities of interest and is conceptually cleaner.
Continuous observation is the method we employ, partly because we also consider it helpful in assessing dependence on starting point, which can be hidden by burn-in.

\paragraph{Multiplicity.}
As is necessary to ensure the desired stationary distribution, we build our ensemble with multiplicity: if we are at state $P$ when a proposal fails, then we count this as a step of the Markov chain and increase the multiplicity of $P$ in our ensemble and in the weighting of the statistics that we gather.

\begin{figure}[htb!]\centering
\includegraphics[width=5.5in]{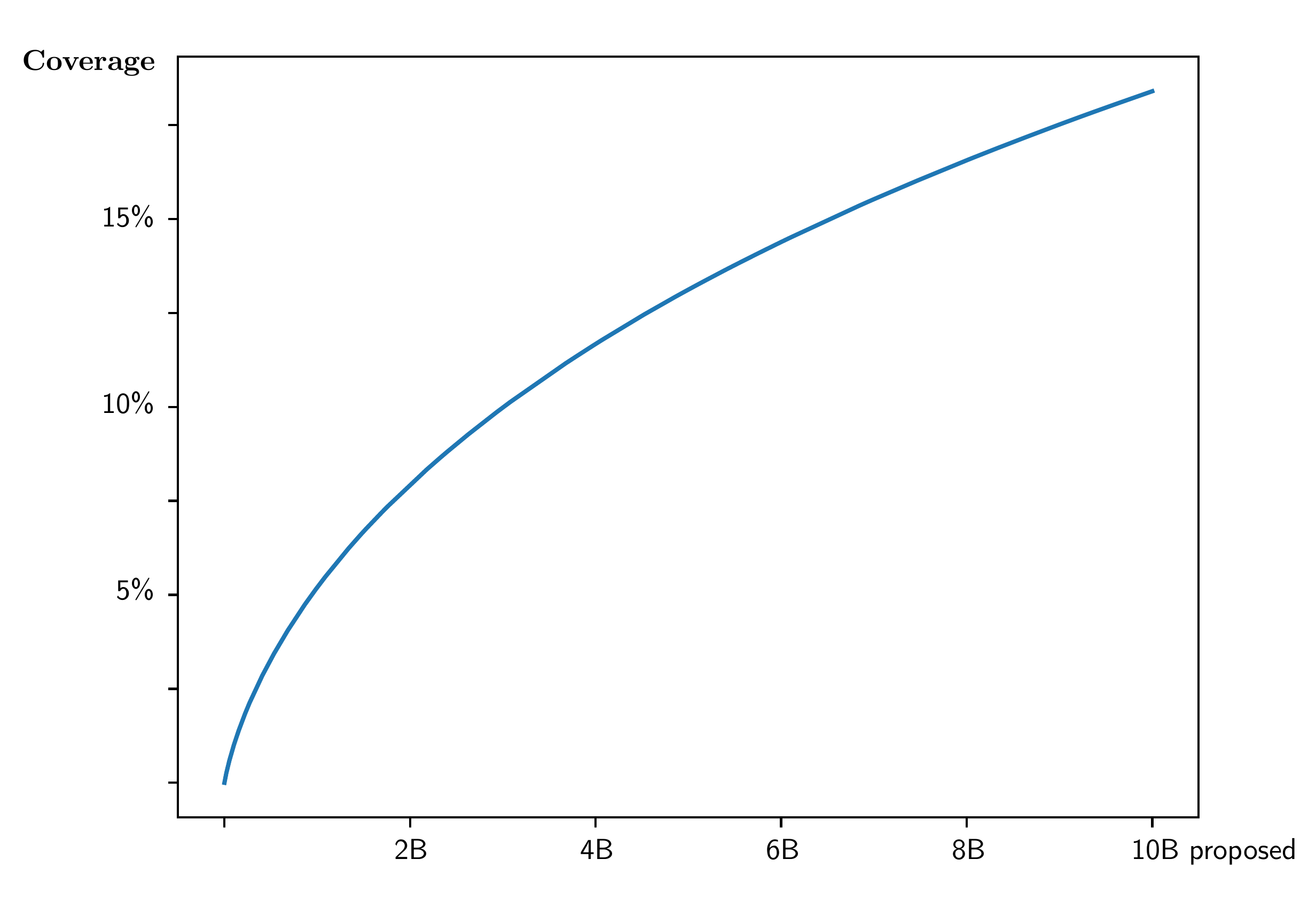}
\caption{{\bf Coverage of the $7 \times 7 \rightarrow 7$ state space.} In ten billion proposed steps on the $7\times 7 \to 7$ state space, this run of \RevReCom has visited roughly 18\% of states, or about 27 million out of the 150 million distinct plans. It is interesting to compare to  Figure~\ref{fig:rrc-convergence}, where the distribution in a summary statistic looks nearly perfect after only 100M proposal steps.}\label{fig:cover}
\end{figure}

\paragraph{Coverage.}
The power of Markov chain methods is that they are capable of producing very accurate estimates while visiting only a tiny portion of a state space.  However, it is also interesting to consider how many distinct plans are encountered; see Figure~\ref{fig:cover}, which considers the $7 \times 7 \rightarrow 7$ state space and shows after ten billion steps only about 18\% of states have been visited. Comparing to Figure~\ref{fig:rrc-convergence}, where we achieved excellent convergence after 100 times fewer steps---and less than 1\% coverage---reminds us that the power of MCMC is to approximate a sampling distribution long before cover time.

\newcommand{\neff}{n_{\textrm{eff}}}

\paragraph{Sample size.}
Separately from issues of whether to continuously observe a chain or subsample, there is a large body of literature focusing on how large of a  sample is needed to reliably estimate statistics \cite{Feller, KarpLubyMadras, Doss}. When samples are independent, if the underlying distribution has mean $\mu$ and variance $\sigma^2$, the average of $n$ samples has variance $\sigma^2/n$.  For statistics that take on values in a bounded range, upper bounds on $\sigma^2$ are easy to find; for however small one would like the variance of the sample average to be, it is then straightforward to determine the needed sample size. 
For a random variable $X$ with expectation $\mu$ and variance $\sigma^2$, by the central limit theorem, the average of $n$ independent random samples of $X$, which we will denote  $X_n$, is asymptotically normal: $X_n \longrightarrow \mathcal{N}(\mu, \sigma^2 / n)$. To ensure that the mean value of some statistic is estimated correctly to one decimal place, one could calculate what is needed for 99\% of the probability mass for $X_n$ fall within a range of of size 0.1.  In a normal distribution, 99\% of the probability mass is between $\mu - 3 SD$ and $\mu + 3SD$, where $SD$ is the standard deviation; ensuring $6 SD < 0.1$ would give the desired property.  For $X_n$, the standard deviation is $\sigma/\sqrt{n}$, and we would want $6 \sigma / \sqrt{n} < 0.1$, or $n > 3600 \sigma^2$.  Estimates of $\sigma^2$, for the original random variable of interest $X$, are required to know how large $n$ should be. For example, if $X$ is the number of Democratic seats in PA in the underlying data from Figure~\ref{fig:PA-compare}, the empirical variance of $X$ is $\sigma^2 = 0.525$ under the Pres16 votes; using Sen16 votes it is $\sigma^2 = 0.551$. Even rounding up to a variance of 1, this suggests that 3600 truly independent random samples should suffice to ensure, with probability 0.99, that the estimate of the mean of $X$ falls within a range of size 0.1. If instead we wanted to target a range of size 0.01, we would need $6 \sigma / \sqrt{n} < 0.01$, or $n > 360,000 \sigma^2$.  
However, this is just for a single statistic, like the mean.
For estimating a statistic other than a mean, one can simply define an auxiliary random variable whose mean is the statistic of interest and apply the same analysis to that random variable. 
If we want to estimate the tails of the distribution, or if we want to make simultaneous estimates for multiple values (as for a district-valued statistic), the needed sample size will go up accordingly, even with perfect access to draws from the desired distribution.
Compounding this problem is that samples are correlated in practice.
When samples are correlated, the average of $n$ random samples may have much higher variance, so the calculations above should be regarded as a lower bound.



\FloatBarrier

\section{Implementation}\label{app:implementation}

\subsection{Implementation of reversible recombination,  with parallelization}
\label{subsec:implementation}

The implementation of \RevReCom used in this paper for the empirical results is written in Rust~\cite{GerryChainRust}. Rust is a programming language increasingly popular in the scientific computing community for its performance and compile-time memory safety guarantees. The implementation uses the general (approximate balance) formulation of \RevReCom given in \S\ref{subsec:approx}, so that the perfect-balance version of the chain given in \S\ref{subsec:exact} is a special case for $\epsilon=0$.

Architecturally, \RevReCom resembles previous  implementations of recombination Markov chains~\cite{GerryChain}, but it is  optimized  to control memory allocation during long runs. Due to these enhancements, the Rust implementation outperforms the principal Python library GerryChain by several orders of magnitude (see Table~\ref{tab:timings}).
Furthermore, the memory safety guarantees in Rust make it well suited to multi-threading. We take advantage of this feature to mitigate the high rejection rate of \RevReCom through a partially parallelized  {\em batching} strategy, which improved the acceptance rate in the benchmarking run to over 300 plans per second.  

The choice of spanning tree is made with Wilson's algorithm, which samples uniformly from possible trees (UST) \cite{wilson}.  For an upper bound $m$ on the number of balance edges, we use $m = 30$ in Virginia and Pennsylvania.  These values are both far larger than the largest number of balance edges we ever observed during an execution of \RevReCom on these states.\footnote{Recall that the resulting distribution in both these cases could differ slightly from $\pi$ because it is conditioned on the process never having observed a tree with more than $m$ $\epsilon$-balance edges. However, observing a large number of balance edges in a single tree is an extremely rare event: even in 2 billion proposals in Pennsylvania, there were at most 23 balance edges at $\epsilon=.01$ (observed 1 time). In Virginia, there were at most 18 balance edges at $\epsilon=.01$ (observed 4 times) in 2 billion proposals.  Because the bound $m$ was set well higher than this, the effect on any conclusions reached because of this subtle conditioning should be negligible.} For the PA and VA runs, the batching was executed with 1024 proposals per batch, split across 8 cores.

\begin{table}[htb!] \centering  
	\begin{tabular}{|c|cc|cccc|c|}

		\hline 
		state & run type & proposals & adjacency & balance & $1/m$ & seam length & accept \\ 
		\hline
		\hline

\hline
\multirow{2}{*}{VA} & \multirow{2}{*}{\ReCom-A} & \multirow{2}{*}{1,000,002} & \multirow{2}{*}{--} & 218,721 &\multirow{2}{*}{--}& \multirow{2}{*}{--} &  \multirow{2}{*}{21.9\%} \\
&&&  & {\em \small 78.1\% rej} &&& \\

\hline
\multirow{2}{*}{VA} & \multirow{2}{*}{\ReCom-B} & \multirow{2}{*}{1,000,005} & 304,053 & 90,950 &\multirow{2}{*}{--}& \multirow{2}{*}{--}& \multirow{2}{*}{9.09\%} \\
&&& {\em \small 69.6\% rej} & {\em \small 70.1\% rej} & & &  \\

\hline
\multirow{2}{*}{VA} & \multirow{2}{*}{\ReCom-C} & \multirow{2}{*}{1,000,005} & \multirow{2}{*}{--} & 225,040 &\multirow{2}{*}{--}& \multirow{2}{*}{--}& \multirow{2}{*}{22.5\%} \\
&&&  & {\em \small 77.5\% rej} & & &  \\

\hline
\multirow{2}{*}{VA} & \multirow{2}{*}{\ReCom-D} & \multirow{2}{*}{1,000,011} & 305,604 & 90,145 &\multirow{2}{*}{--}& \multirow{2}{*}{--}& \multirow{2}{*}{9.01\%} \\
&&& {\em \small 69.4\% rej} & {\em \small 70.5\% rej} & & & \\

\hline
\multirow{2}{*}{VA} & \multirow{2}{*}{\RevReCom} & \multirow{2}{*}{1,000,902} & 301,205 & 76,009 & 2524 & 578 & \multirow{2}{*}{0.0577\%} \\
&&& {\em \small 69.9\% rej} & {\em \small 74.8\% rej} &  {\em \small 96.7\% rej} &  {\em \small 77.1\% rej} &  \\

\hline
\hline

\hline
\multirow{2}{*}{PA} & \multirow{2}{*}{\ReCom-A} & \multirow{2}{*}{1,000,001} & \multirow{2}{*}{--} & 246,928 & \multirow{2}{*}{--} & \multirow{2}{*}{--}& \multirow{2}{*}{24.7\%} \\
&&&  & {\em \small 75.3\% rej} & & &  \\

\hline
\multirow{2}{*}{PA} & \multirow{2}{*}{\ReCom-B} & \multirow{2}{*}{1,000,021} & 224,804 & 73,159 & \multirow{2}{*}{--} & \multirow{2}{*}{--}& \multirow{2}{*}{7.32\% } \\
&&& {\em \small 77.5\% rej} & {\em \small 67.5\% rej} &  & & \\

\hline
\multirow{2}{*}{PA} & \multirow{2}{*}{\ReCom-C} & \multirow{2}{*}{1,000,002} & \multirow{2}{*}{--} & 250,931 & \multirow{2}{*}{--}  & \multirow{2}{*}{--}& \multirow{2}{*}{25.1\%}\\
&&&  & {\em \small 74.9\% rej} & & &  \\

\hline
\multirow{2}{*}{PA} & \multirow{2}{*}{\ReCom-D} & \multirow{2}{*}{1,000,042} & 227,172 & 72,225 &\multirow{2}{*}{--}& \multirow{2}{*}{--}& \multirow{2}{*}{7.22\%}\\
&&& {\em \small 77.3\% rej} & {\em \small 68.2\% rej} & & &  \\

\hline
\multirow{2}{*}{PA} & \multirow{2}{*}{\RevReCom} & \multirow{2}{*}{1,000,482} & 217,050 & 66,949 & 2174 & 443  &  \multirow{2}{*}{0.0443\%}\\
&&& {\em \small 78.3\% rej} & {\em \small 69.2\% rej} & {\em \small 96.8\% rej} &  {\em \small 79.6\% rej}&\\

\hline
\hline

\multirow{2}{*}{7x7} & \multirow{2}{*}{\ReCom-A} & \multirow{2}{*}{1,000,001} & \multirow{2}{*}{--} & 532,692 &\multirow{2}{*}{--}& \multirow{2}{*}{--}&  \multirow{2}{*}{53.3\%}\\
&&&  & {\em \small 46.7\% rej} & & & \\

\hline
\multirow{2}{*}{7x7} & \multirow{2}{*}{\ReCom-B} & \multirow{2}{*}{1,000,004} & 467,867 & 289,705 &\multirow{2}{*}{--}& \multirow{2}{*}{--}& \multirow{2}{*}{29\%}\\
&&& {\em \small 53.2\% rej} & {\em \small 38.1\% rej} & & &  \\

\hline
\multirow{2}{*}{7x7} & \multirow{2}{*}{\ReCom-C} & \multirow{2}{*}{1,000,001} &\multirow{2}{*}{--} & 542,793 &\multirow{2}{*}{--}& \multirow{2}{*}{--}& \multirow{2}{*}{54.3\%}\\
&&& & {\em \small 45.7\% rej} & & &  \\

\hline
\multirow{2}{*}{7x7} & \multirow{2}{*}{\ReCom-D} & \multirow{2}{*}{1,000,001} & 469,550 & 293,290 &\multirow{2}{*}{--}& \multirow{2}{*}{--}& \multirow{2}{*}{29.3\%}\\
&&& {\em \small 53.0\% rej} & {\em \small 37.5\% rej} & & &  \\

\hline
\multirow{2}{*}{7x7} & \multirow{2}{*}{\RevReCom} & \multirow{2}{*}{1,000,008} & 465,932 & 292,039 & \multirow{2}{*}{--}& 164,252 & \multirow{2}{*}{16.4\%}\\
&&& {\em \small 53.4\% rej} & {\em \small 37.3\% rej} && {\em \small 43.8\% rej} &  \\

\hline


	   \end{tabular}
        \caption{{\bf Acceptance rates}. Run statistics, showing the number of proposals accepted at each stage, for recombination runs on Virginia (precincts, $k=11$, $\epsilon=.01$, $m=30$), Pennsylvania (VTDs, $k=18$, $\epsilon=.01$, $m=30$), and a $7\times 7$ grid ($k=7, \epsilon=0$, $m=1$).  Roughly 1 million steps are initially proposed in each trial; irregular numbers are due to the batching technique, as described in \S\ref{subsec:implementation}. 
        The $1/m$ column is estimated with a geometric variable since it is integrated in the seam length step, as implemented.
        One observation is that the variants using MST (A and C) are less likely to find a balance edge than the others, which use UST. This is expected behavior, since MST trees are weighted toward having higher-degree vertices, making them harder to split in a balanced fashion.}
\label{tab:rejection}
\end{table}

We sketch the batching strategy here.
The low acceptance rate of \RevReCom on real-world graphs (see Table~\ref{tab:rejection}) enables us to secure  performance gains by using multiple cores. As a general matter, Markov chains resist parallelization because, by definition, the next step is probabilistically determined by the current state, so a single random walker must consider the proposals at each step.  However, if the great majority of proposals are being rejected in a given chain, then we can get an efficiency boost by using parallel workers  at the proposal stage that collectively consider a few hundred proposals simultaneously.  They can then be ordered randomly and the accepted proposal with lowest index can be passed back to the main thread, at a potentially significant time savings. This is an instance of what is sometimes called ``speculative execution''---work is performed that may not be used, in order to minimize lag time.  We have found work by Brockwell from 2006 proposing a similar batching strategy, where he uses the term ``pre-fetching'' for the partial parallelization \cite{Brockwell}. 
The batch size should be large enough for the advantages to surpass the cost of synchronization overheads, but without leading to many wasted samples.
Empirically, we find that a batching strategy is effective for reducing wall-clock compute time on full-scale redistricting problems.

Python GerryChain loads shapefiles or graphs as its input format; the Rust implementation loads graphs with integer-valued attributes. 
Very long \RevReCom runs can save selected summary statistics straight to JSONL files. Generally, outputs are available in various serialization formats, including a highly compressed file format called {\sf BEN} (\href{https://gerrytools.readthedocs.io/en/latest/user/ben/}{\tt gerrytools.readthedocs.io/en/latest/user/ben}).

\subsection{Testing algorithms of other authors}\label{app:other-authors}

To test Forest \ReCom and SMC, we rely on the implementations provided by their originators, found at \cite{Forest-repos} and \cite{Redist-repo}.  
Forest \ReCom is written in Python and Julia while SMC uses {\sf R} and C++.  
(We note that other implementations exist as well; for instance, the Redist repo that is the primary home for SMC also contains an {\sf R} implementation of Forest \ReCom.)
Each codebase has its own ways of loading data and storing outputs, but our replication repo includes helper functions to make inputs and outputs interoperable.
Replication materials can be found in \cite{replication}.

SMC keeps a great deal of graph data in memory during a run, causing significant RAM consumption (see Table~\ref{tab:timings}); only after a batch run is done are plans written to disk.

We have made a serious effort to collect and present samples by each of the methods in the manner intended by the authors.  Typically, Markov chain authors use multistart and enlargement tests  to provide convergence heuristics (i.e., comparing runs from multiple starting points, and checking to see that much longer runs perform similarly to the reported runs).
This is what we do here, rather than using subsampling and/or the smallest acceptable ensemble sizes to compete for the appearance of efficiency.  
(We have confirmed that the use of burn-in and subsampling would not materially alter any findings we present.)
For Markov chain methods, there is a natural rule of thumb available to choose a sample size for a given accuracy target.  A user who desires a sample within $\dwass<\tau$ of the stationary distribution, for some threshold $\tau$, should at a minimum run chains from different starting conditions until the time-$t$ ensembles are within $2\tau$ of each other.  

The SMC authors advocate for the use of the Gelman-Rubin convergence diagnostic $\widehat{R}$, which compares within-sample variance to between-sample variance, usually referring to ``our heuristic convergence check that $\widehat{R}\le 1.05$" \cite{SMC}.  Following this scheme, if it tends to be the case that two samples with $S=S_0$ plans pass this test, then we treat $S_0$ as an adequate batch size to get an accurate sample.
Information on the $\widehat{R}$ diagnostic can be found at \cite{Gelman1992Inference,Vehtari2021Rank}.

\subsection{Timing comparisons}

\begin{table}[htb!] \centering 
\begin{tabular}{lllll}
{\bf MCMC Method} & {\bf Notes} & {\bf Accepted} & {\bf Time} & {\bf Rate}\\
\hline 
Python \ReCom-A & 1 core & 1000 plans & 51.56 sec & 19.4 accepted/sec\\
Rust \ReCom-A & 1 core & 485,987 plans & 107.79 sec & 4509 accepted/sec\\
Rust \ReCom-A & 4 cores & 485,983 plans & 71.08 sec & 6837 accepted/sec\\
Rust \ReCom-A & 4 cores & 487,239 plans & 89.68 sec & 5443 accepted/sec\\

Rust \RevReCom & 1 core & 2136 plans & 43.791 sec & 48.8 accepted/sec\\
Rust \RevReCom & 4 cores, no batching & 2033 plans & 25.537  sec & 79.6 accepted/sec\\
Rust \RevReCom & 10 cores, batches of 12 & 2022 plans & 5.928 sec & 341.1 accepted/sec\\
Rust \RevReCom & 10 cores, batches of 16 & 1715 plans & 5.401 sec & 317.5 accepted/sec\\
Rust \RevReCom & 4 cores, batches of 32 & 2105 plans & 9.331 sec & 225.6 accepted/sec\\
Forest \ReCom & & 2259 plans & 682.24 sec & 3.3 accepted/sec 
\end{tabular}
\end{table}

\begin{table}[htb!]\centering 
\begin{tabular}{llll}
 {\bf SMC batch size ($S$)}& {\bf Memory (RAM)} & {\bf Time} & {\bf Rate}\\
\hline 
 5000 plans& 0.67 GB & 
107 sec & 46.7 produced/sec \\
 20,000 plans& 1.76 GB & 
354 sec & 56.5 produced/sec \\
 100,000 plans& 10.54 GB & 
1947 sec & 51.4 produced/sec 
\end{tabular}
\caption{{\bf Basic timing comparison for MCMC methods and SMC.}
We include selected timing information to facilitate apples-to-apples comparisons between implementations.  Timings are variable, so these single-run figures are intended to be illustrative.
We emphasize that speed is not related to sample quality except that practical applications will need runtimes to be on the scale of hours or days, not weeks.
For all benchmarking runs, we use the Virginia precinct dual graph with the CD12 seed (Congressional districts from 2012), $\epsilon=.05$, and minimal updaters. We set $m = 30$ for \RevReCom and SMC.
The runs were conducted on an Apple M1 Pro laptop (10 cores) with 16 GB RAM.  Python runs were executed in version 3.10.2.
SMC runs performed in a Linux virtual machine with 5 cores on the same hardware.}
\label{tab:timings}
\end{table}

\FloatBarrier
\section{Empirics}\label{app:empirics}

\subsection{Data and methods for comparisons on Pennsylvania, Virginia, and the $7\times 7$ grid}

\paragraph{Data.} The 158,753,814 configurations in the $7\times 7\to 7$ districting problem were found using a tool called {\sf enumpart} that was created by Kosuke Imai's research team and can be found in the same Redist repo as the SMC code \cite{Redist-repo}. 
The plans were then re-weighted by their spanning tree counts to produce the dataset used in \S\ref{sec:7x7}.  Supporting code is available in the replication repo for this paper \cite{replication}.  We note that the $7\times 7\to 7$ state space was confirmed to be connected by a brute-force computational check.

The maps assessed above in Pennsylvania include the Legislative proposals from 2011 and 2018 and the court's Remedial plan from 2018.  
The ``8th Grade Class" map was created by students of Jon Kimmel at Westtown School in Chester, PA, and was covered by the local press ({\em How difficult is it to redistrict Pennsylvania? ‘Not very,’ say area eighth-graders}, available from \href{https://whyy.org/segments/difficult-redistrict-pennsylvania-not-say-area-eighth-graders/}{\tt whyy.org}).
In Virginia, we have used the enacted Congressional maps from 2012 and 2016 as seed plans in Figure~\ref{fig:VA}.
In addition, precinct-level Presidential returns from 2016 and U.S. Senate returns from 2016 were used to study partisan outcomes across ensembles of alternative plans.  Generally both plans and election data for these time periods are publicly available from Redistricting Data Hub (\href{https://redistrictingdatahub.org/}{\tt redistrictingdatahub.org}).
The specific electoral datasets used here were cleaned and prepared by the members of the MGGG Redistricting Lab for research use and are also shared in \cite{replication}.

\paragraph{Methods: Choice of election data.} 
For Figure~\ref{fig:PA-compare} (top), we use the vote counts by precinct for Clinton (Democratic) and Trump (Republican), and for McGinty (Democratic) and Toomey (Republican)---the major-party candidates in the 2016 contests for President and U.S. Senate, respectively.
The number of Democratic seats is simply the count of districts in which the Democrat received more votes than the Republican, obtained by summing over the precincts.  We note that this is a standard approach: we use statewide (exogenous) elections rather than Congressional (endogenous) elections for the underlying vote pattern in an ensemble analysis, so that moving the district lines does not run up against issues from uncontested districts, variable incumbency advantage, and other district-specific confounding variables.
The use of statewide elections in serial rather than composite ensures that the D/R votes draw from ``naturalistically" observed patterns; one can still interpret individual contests and be mindful of their anomalies.
Though it is standard, this approach does have detractors in political science, who prefer to use regression techniques to blend endogenous races together into a synthetic election while controlling for various confounding factors.

\clearpage
\subsection{Detailed SMC diagnostics}

\begin{figure}[htb!]\centering
\includegraphics[width=6in]{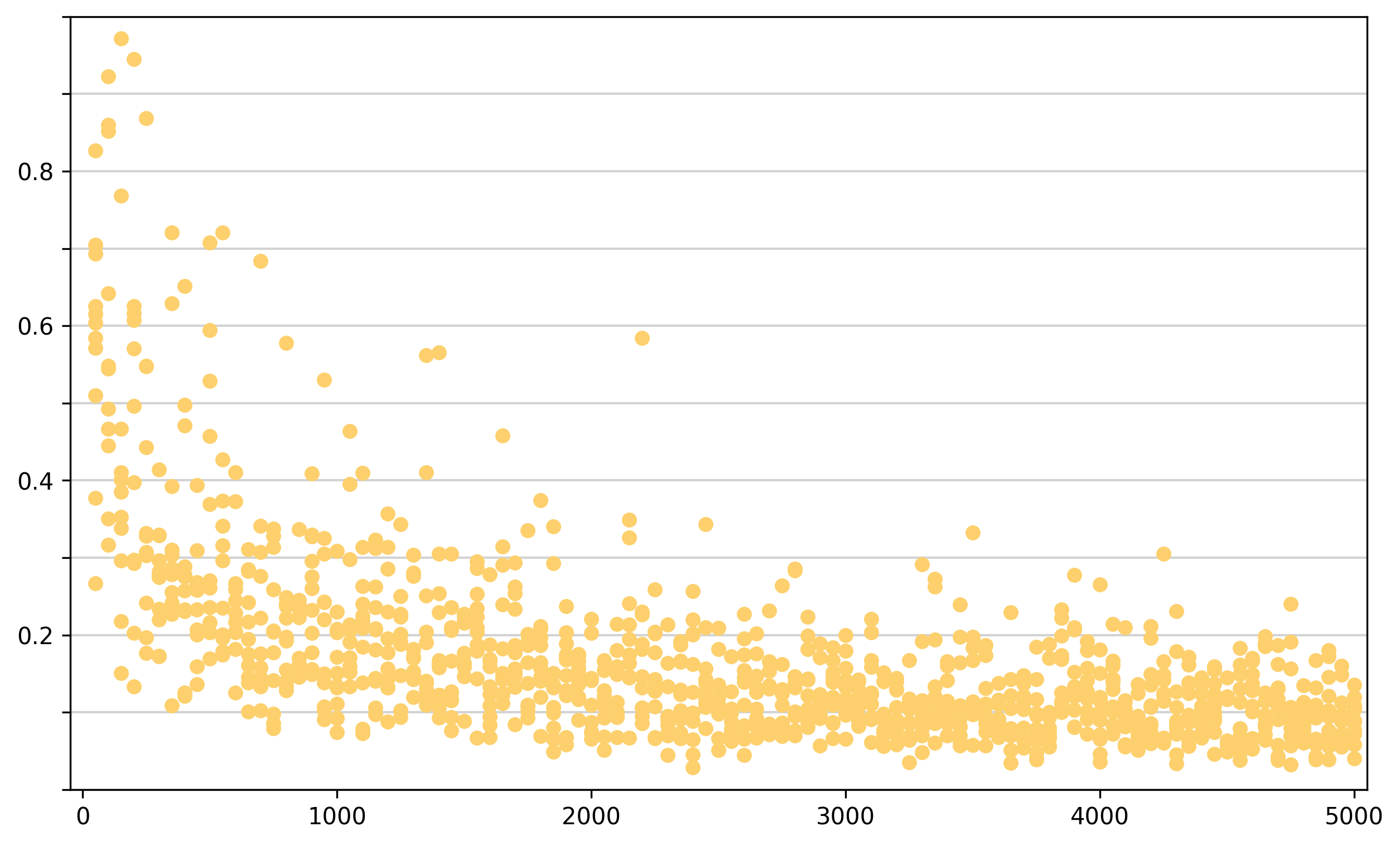}

\vspace{30pt}

\begin{tabular}{c|llllllllllll}
$S$& 50 &100&150 &200&250 &300&350 &400&450 &500&550&600\\
\hline 
{\small \# pairs w/ } &\multirow{3}{*}{40}&
\multirow{3}{*}{37}&\multirow{3}{*}{43}&\multirow{3}{*}{37}&\multirow{3}{*}{48}&\multirow{3}{*}{54}&\multirow{3}{*}{49}&\multirow{3}{*}{54}&\multirow{3}{*}{55}&\multirow{3}{*}{53}&\multirow{3}{*}{54}&\multirow{3}{*}{55}\\
$\widehat{R}<1.05$ & \\
{\small out of 55}& 
\end{tabular}

\caption{{\bf Detailed SMC diagnostics.}
TOP: The scatterplot shows the Wasserstein distance of the sampled cut edge distribution from the  $\pi$-weighted ground truth for small batches of SMC plans for the $7\times 7$ problem ($S=50,100,150,\dots,5000$).
Batch sizes up to $S=100,000$ are shown in Figure~\ref{fig:7x7}.
BOTTOM: For 11 runs at each each $S$ value, we calculate $\widehat{R}$ for the cut edges statistic for  all ${11 \choose 2} = 55$ pairs of runs and count how many of them pass the $\widehat{R} < 1.05$ test pairwise.}
\label{fig:SMC-small}
\end{figure}

The SMC authors use the Gelman--Rubin $\widehat{R}$ statistic, as their main convergence diagnostic. Though better known for use with MCMC methods, it compares within-sample to between-sample variability, so can be used in this setting.  More unusually, they designate a particular threshold of $\widehat{R}\le 1.05$ for practical use in redistricting problems, as cited in \S\ref{app:other-authors}.  The statistics shown in Figure~\ref{fig:SMC-small} illustrate that a sample size of $S=400$ would be deemed adequately converged by this standard.
This produces accuracy similar to the heuristically targeted methods \ReCom-A,B,C in this problem (see Figure~\ref{fig:7x7}).  

\clearpage
\subsection{$50\times 50$ experiments}

Finally, we compare \ReCom-A,B,C,D, \RevReCom, and SMC samples for the problem of dividing a $50\times 50$ grid into 10, 25, and 50 districts.  

\begin{figure}[htb!] \centering 
\begin{tikzpicture}
\node at (0,4.5) {\includegraphics[width=4.4in]{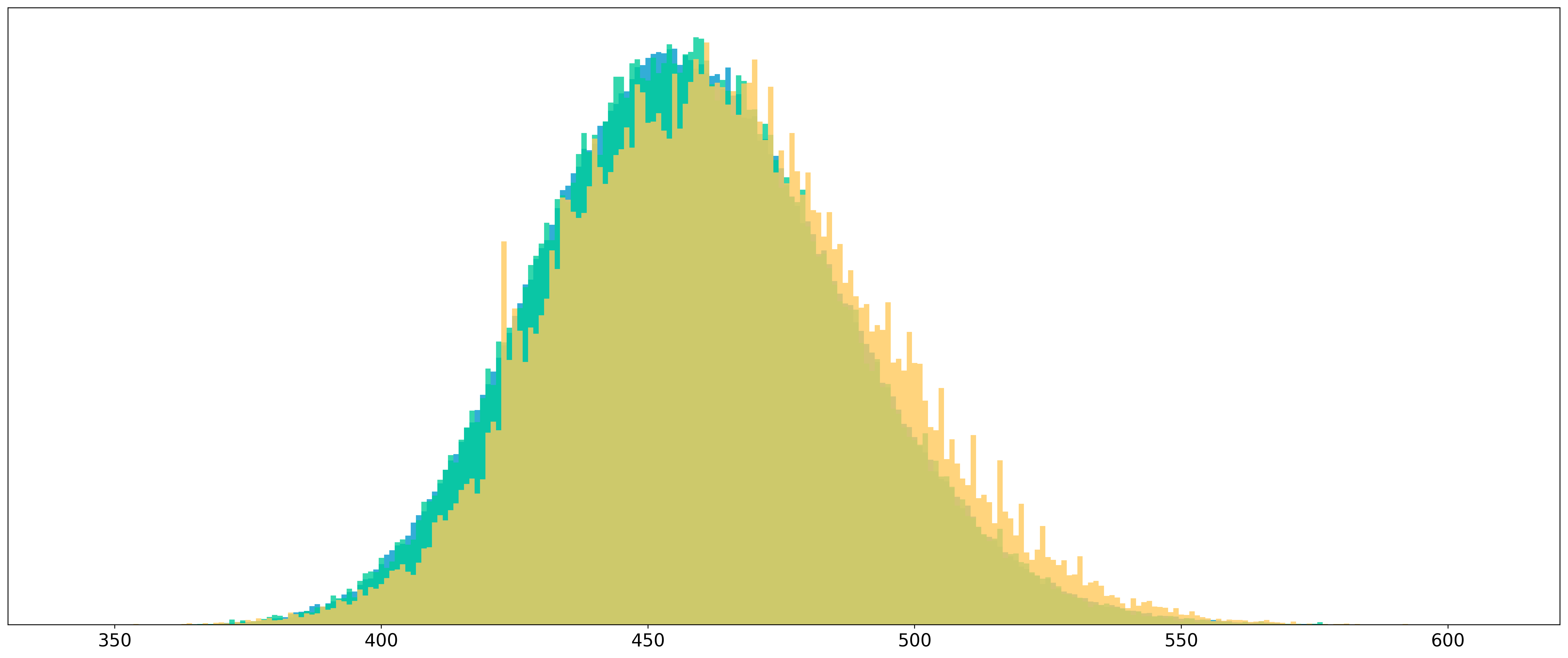}};
\node at (4,5.6) {\includegraphics[width=80pt]{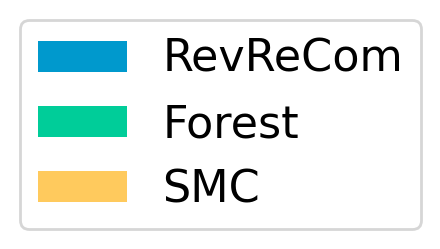}};
\node at (0,0) {\includegraphics[width=4.4in]{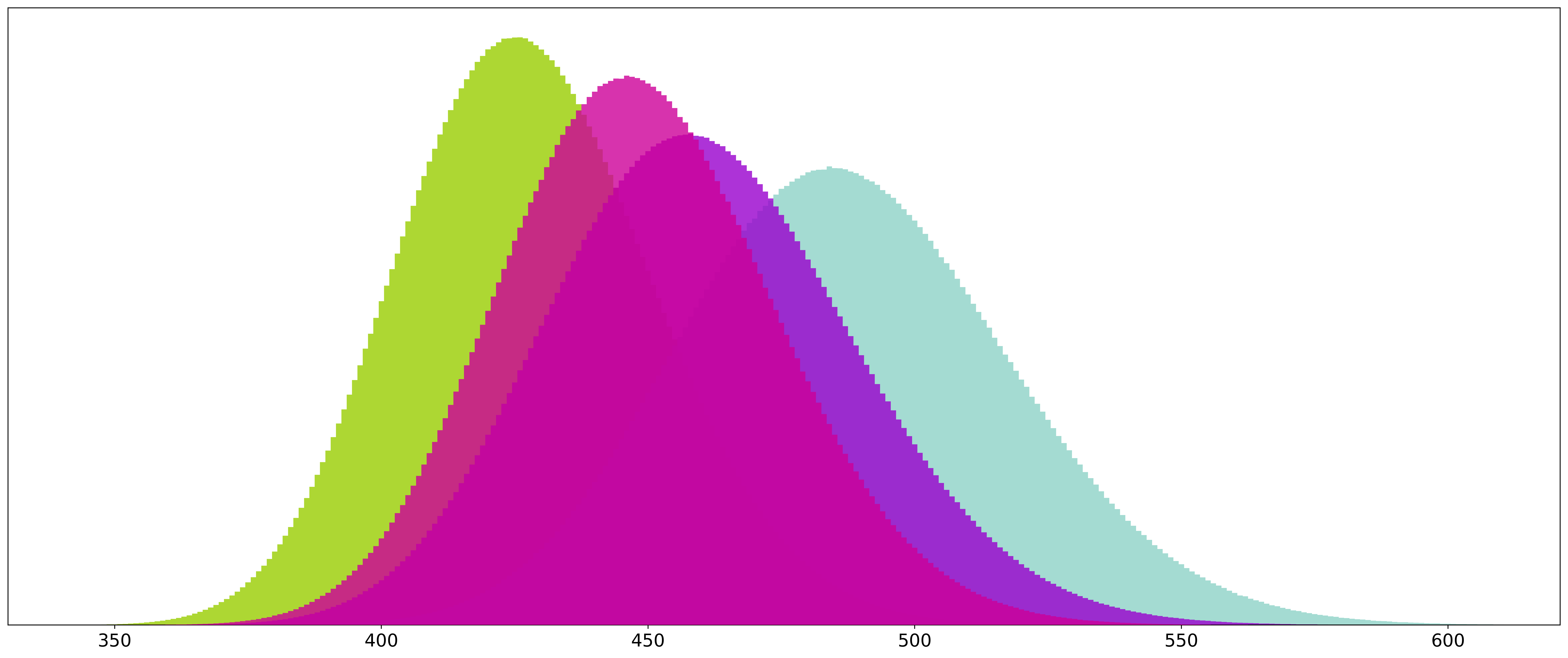}};
\node at (4,1.1) {\includegraphics[width=72pt]{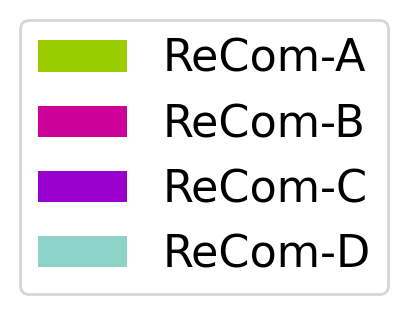}};

\node at (0,7.1) {\Large $\mathbf{50\times 50 \to 10}$};
\end{tikzpicture}

\vspace{3pt}

\begin{tikzpicture}
\node at (0,4.5) {\includegraphics[width=4.4in]{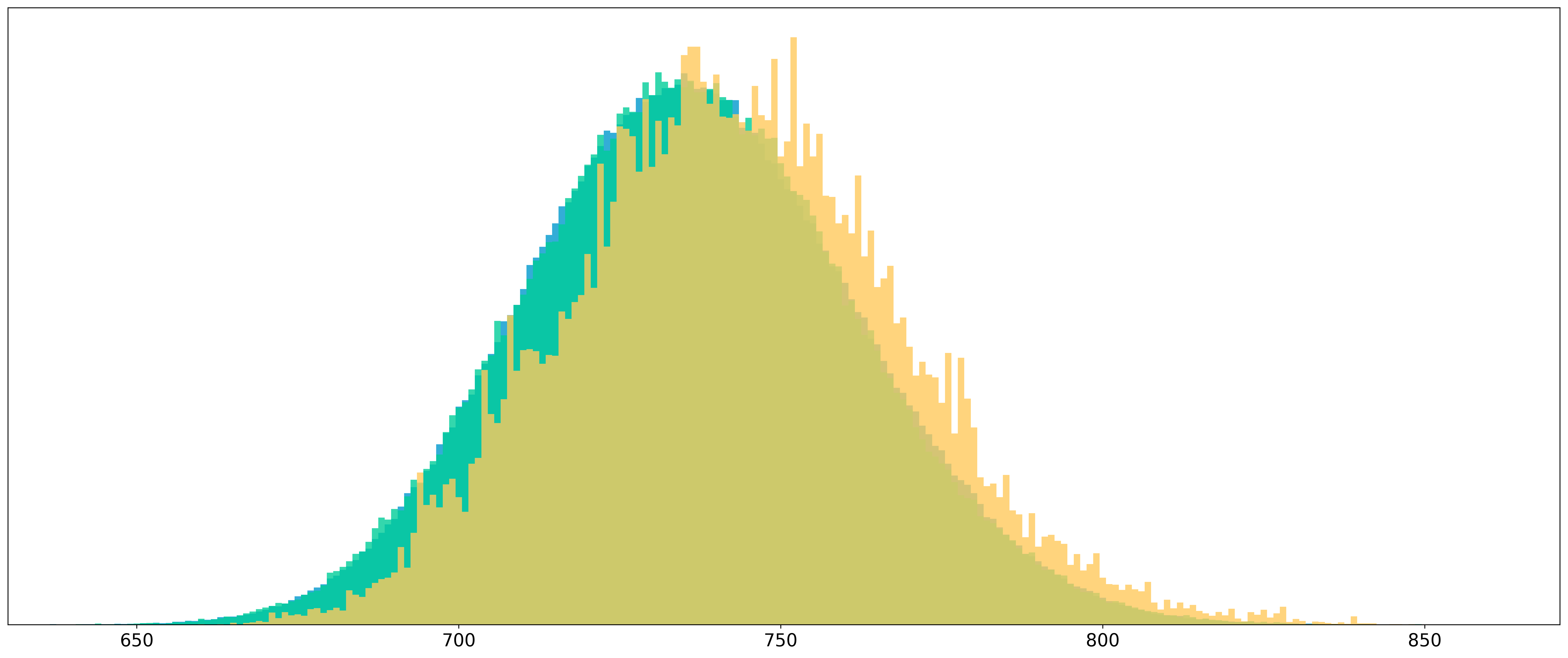}};
\node at (4,5.6) {\includegraphics[width=80pt]{figSIREV/50x50_10_dist_forest_rrc_smc_comparison_legend.png}};
\node at (0,0) {\includegraphics[width=4.4in]{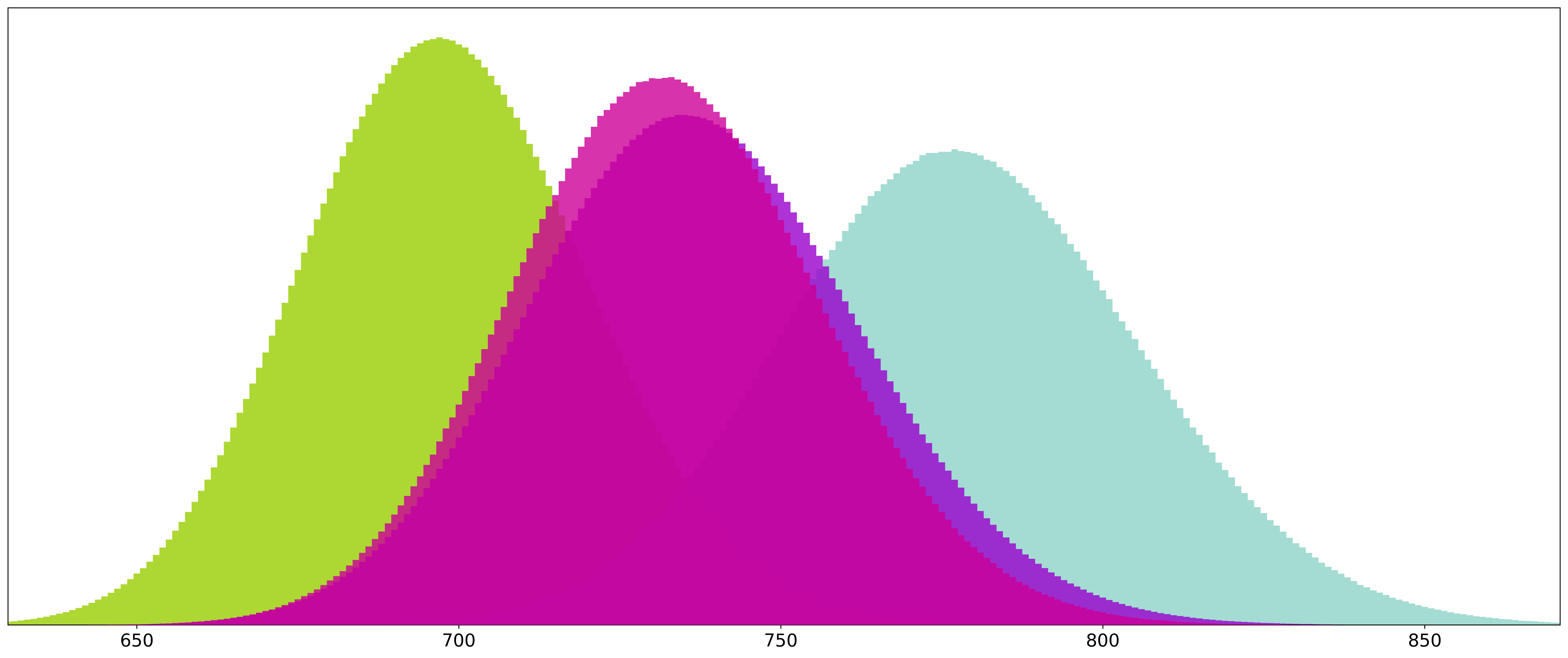}};
\node at (4,1.1) {\includegraphics[width=72pt]{figSIREV/50x50_10_dist_ReCom_comparison_legend.png}};
\node at (0,7.1) {\Large $\mathbf{50\times 50 \to 25}$};
\end{tikzpicture}
\end{figure}

\begin{figure}[htb!] \centering 
\begin{tikzpicture}
\node at (0,6.4) {\includegraphics[width=6in]{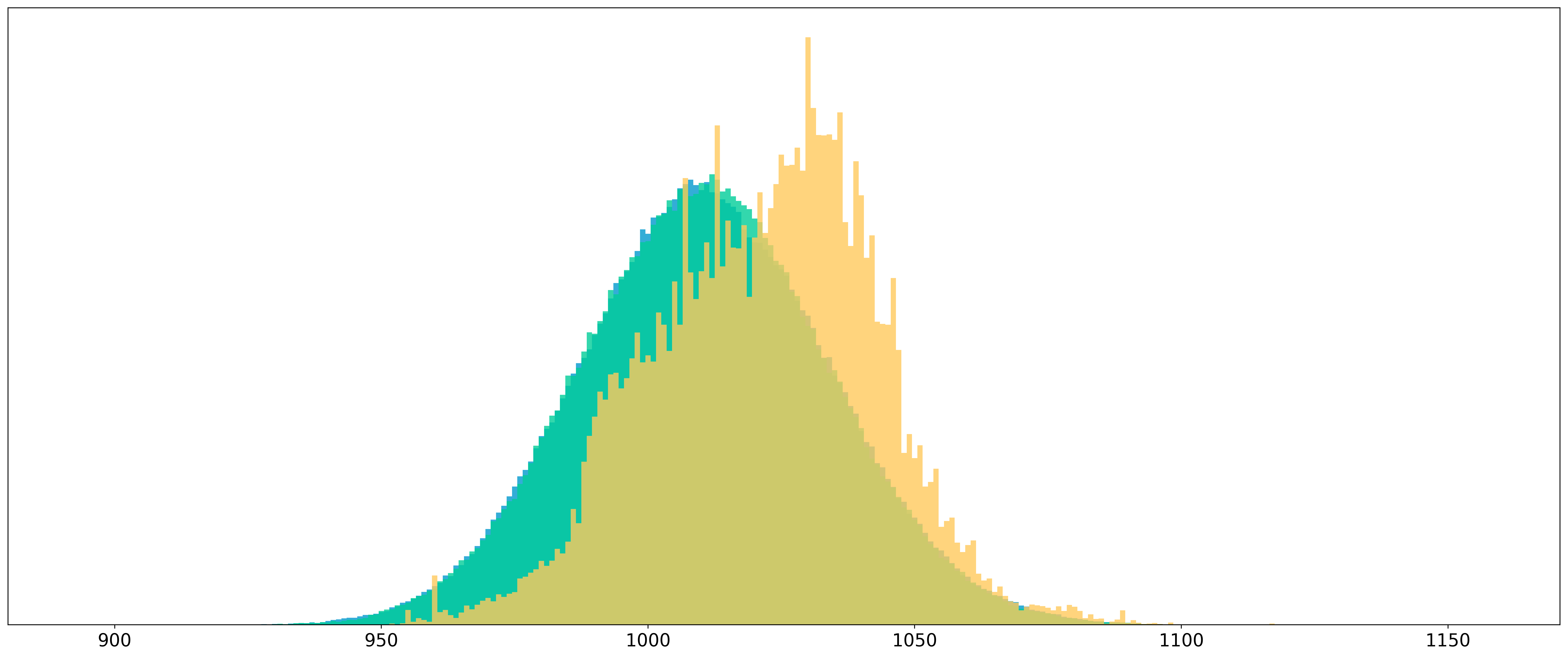}};
\node at (5.8,6.2+1.8) {\includegraphics[width=80pt]{figSIREV/50x50_10_dist_forest_rrc_smc_comparison_legend.png}};
\node at (0,0) {\includegraphics[width=6in]{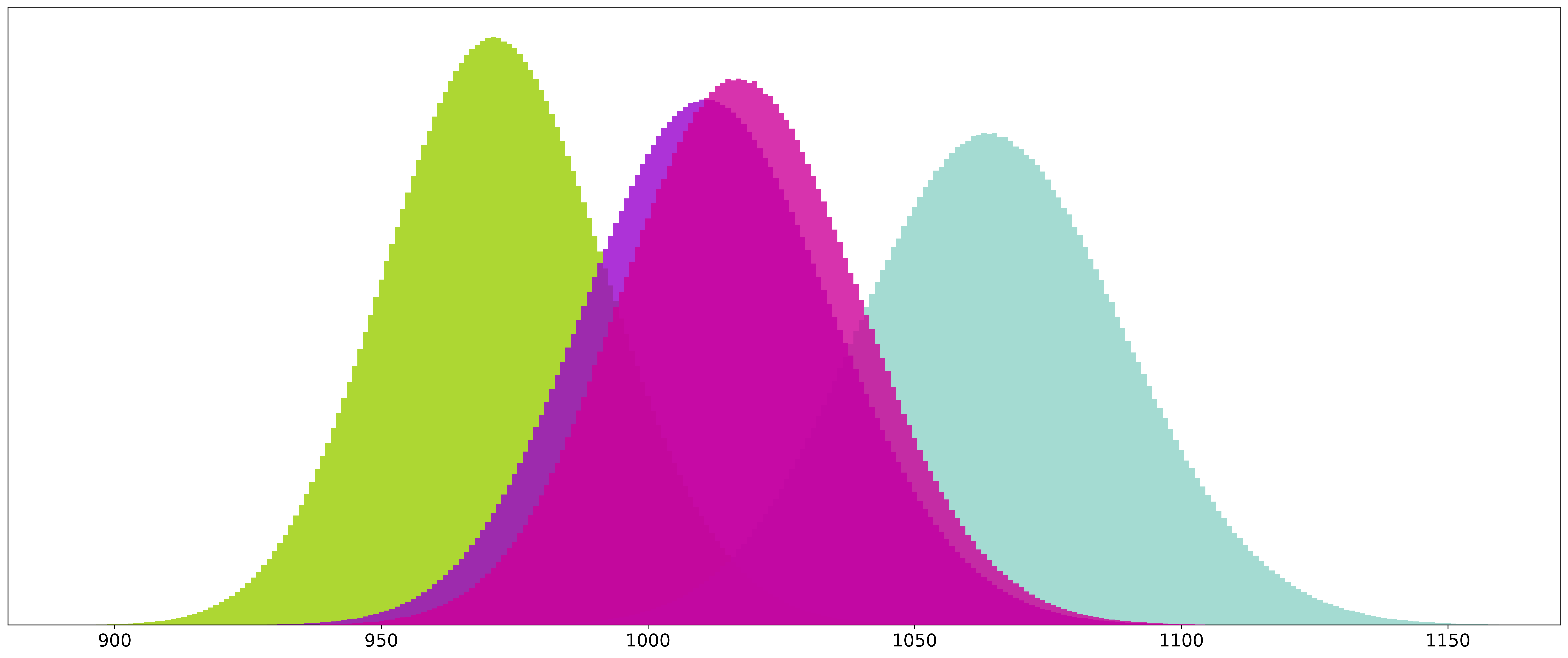}};
\node at (5.8,1.8) {\includegraphics[width=72pt]{figSIREV/50x50_10_dist_ReCom_comparison_legend.png}};
\node at (0,10) {\Large $\mathbf{50\times 50 \to 50}$};
\end{tikzpicture}

\caption{{\bf $50\times 50$ grid comparisons.} }
        \label{fig:50x50-2}
\end{figure}

\FloatBarrier
All methods are shown at their largest practical sample sizes (see \S\ref{sec:samplers}).
The three methods that provably target the spanning tree distribution $\pi$ are shown as the top plot in each pair, and the Markov chain methods (\RevReCom and Forest \ReCom) give visually indistinguishable results.  
Since SMC sample quality improves with batch size, we use $\pi$-targeted runs with $S=100,000$ for the best possible results.  (Published SMC results usually rely on much smaller batch sizes, even for state-level redistricting.)  
At this size, the SMC technique is able to get a  sample that reasonably resembles the recombination methods when dividing the $50\times 50$ grid into $k=10$ districts, but its performance suffers, even pushed to the largest achievable sample size, with more districts.   The performance of the recombination samplers does not show signs of degrading as the number of districts increases.    We note that the size of this test problem---2500 units of a $50\times 50$ grid, and ten to 50 districts---is realistic for real-world settings. For instance, Pennsylvania has about 9000 precincts, $k=18$ or now $k=17$ Congressional districts, $k=50$ state Senate districts, and $k=203$ state House districts.  In Virginia, there are about 5000 precincts, and the number of districts is $k=11,40,100$, respectively.

Consistent with the discussion throughout this paper, the four methods \ReCom-A,B,C,D all look extremely well converged, but to a steady state that is not identical with $\pi$.  \ReCom-C resembles $\pi$ closely in each example here.  Finally, we remark that, going beyond convergence considerations, large samples will still be needed in all methods for estimating finely-binned or high-dimensional statistics.

\end{document}